\documentclass[journal]{IEEEtran}
\newcommand{\ee}{\epsilon}			
\usepackage{epsf,psfrag,amssymb,amsfonts,color,cite,fancybox}
\usepackage[mathscr]{eucal}
\usepackage[dvips]{graphicx}
\usepackage{ifpdf}
\usepackage{longtable}	
\usepackage{multicol}
\usepackage{float}

\ifCLASSINFOpdf

\else

\fi

\usepackage[cmex10]{amsmath}
\usepackage[caption=false,font=footnotesize]{subfig}

\begin{document}
\renewcommand{\thefigure}{\arabic{figure}}
\renewcommand{\thesubfigure}{\alph{subfigure}}

\title{Analytical Studies of Quasi Steady-State Model in Power System Long-Term Stability Analysis}

\author{Xiaozhe Wang,~\IEEEmembership{Student Member,~IEEE,}
        Hsiao-Dong Chiang,~\IEEEmembership{Fellow,~IEEE.}
\thanks{This work was supported by the Consortium for Electric Reliability
Technology Solutions provided by U.S. Department No. DE-FC26-
09NT43321.}
\thanks{Xiaozhe Wang is with the School of Electrical and Computer Engineering, Cornell University, Ithaca,
NY 14853 USA, email: xw264@cornell.edu.}
\thanks{Hsiao-Dong Chiang is with the School of Electrical and Computer Engineering, Cornell University, Ithaca, NY 14853 USA, email:hc63@cornell.edu.}
\thanks{Copyright (c) 2013 IEEE. Personal use of this material is
permitted. However, permission to use this material
for any other purposes must be obtained from the IEEE by
sending an email to pubs-permissions@ieee.org.}%
}

\maketitle

\begin{abstract}
In this paper, a theoretical foundation for the Quasi Steady-State (QSS) model in power system long-term stability analysis is developed. Sufficient conditions under which the QSS model gives accurate approximations of the long-term stability model in terms of trajectory and $\omega$-limit set are derived. These sufficient conditions provide some physical insights regarding the reason for the failure of the QSS model. Additionally, several numerical examples are presented to illustrate the analytical results derived.
\end{abstract}


\IEEEpeerreviewmaketitle

\section{Introduction}
\IEEEPARstart{T}HE ever-increasing loading of transmission networks together with a steady increase in load demands has pushed many power systems ever closer to their stability limit \cite{Chiang:book}-\cite{SauerPai:book}. Long-term stability has become more and more important for secure operation of power systems. However, the long-term stability model is large and involves different time scales. The time domain simulation approach for the long-term stability model is expensive in terms of computational efforts and data processing. These constraints are even more stringent in the context of on-line stability assessment.
The quasi steady-state (QSS) proposed in \cite{Cutsem:book}-\cite{Cutsem:artical} tried to reach a good compromise between accuracy and efficiency for long-term stability analysis. The assumptions behind the QSS model that the post-fault transient stability model is stable and the long-term stability model is singularity-free are not necessarily true. There have been some efforts attending to address these issues \cite{Cutsem:artical2}-\cite{Wang:artical}. However, less attention has been paid to another critical issue that even these assumptions are satisfied, the QSS model may still provide incorrect approximations for the long-term stability model. Some counter examples in which the QSS model were stable while the long-term stability model underwent long-term instabilities were presented in \cite{Wangxz:article}. Since the QSS model can not consistently provide correct stability analysis of the long-term stability model, there is a great need to identify conditions under which the QSS model works.
In this paper, sufficient conditions under which the QSS model can provide correct approximations for the long-term stability model are developed. Briefly speaking, if neither the long-term stability model nor the QSS model meets a singularity, then the QSS model provides correct approximations for the long-term stability model in terms of trajectory if the QSS model moves along the stable component of its constraint manifold and the projection of each point on the trajectory of the long-term stability model lies inside the stability region of the corresponding transient stability model. Moreover, if the QSS model converges to a long-term stable equilibrium point (SEP), then the long-term stability model will converge to the same point. Several numerical examples in which the QSS model succeeded or failed are analyzed by the derived analytical results.

This paper is organized as follows. Section \ref{sectioncompleteqssmodel} recalls basic concepts of power system models, and Section \ref{mathprelim} introduces mathematical preliminaries in nonlinear system theories. Then sufficient conditions of the QSS model are derived in Section \ref{analyticalstudies}, and several numerical examples are analyzed based on the derived theorems in Section \ref{numericalstudies}. Conclusions and perspectives are stated in Section \ref{conclusion}.

\section{Power System Models}\label{sectioncompleteqssmodel}
The long-term stability model, or interchangeably complete dynamic model, for calculating system dynamic response relative to a disturbance can be described as:
\begin{eqnarray}
\dot{z}_{c}&=&\ee{h}_c({z_c,z_d,x,y})\label{slow ode}\\
{z}_d(k+1)&=&{h}_d({z_c,z_d(k),x,y})\label{slow dde}\\
\dot{{x}}&=&{f}({z_c,z_d,x,y})\label{fast ode}\\
{0}&=&{g}({z_c,z_d,x,y})\label{algebraic eqn}
\end{eqnarray}

Equation (\ref{algebraic eqn}) describes the electrical transmission system and the internal static behaviors of passive devices, and (\ref{fast ode}) describes the internal dynamics of devices such as generators, their associated control systems, certain loads, and other dynamically modeled components. ${f}$ and ${g}$ are continuous functions, and vector ${x}$ and ${y}$ are the corresponding short-term state variables and algebraic variables. Besides, Equations (\ref{slow ode}) and (\ref{slow dde}) describe long-term dynamics including exponential recovery load, turbine governor, load tap changer (LTC), over excitation limiter (OXL), etc. ${z}_c$ and ${z}_d$ are the continuous and discrete long-term state variables respectively, and $1/\ee$ is the maximum time constant among devices. Since transient dynamics have much smaller time constants compared with those of long-term dynamics, $z_c$ and $z_d$ are also termed as slow state variables, and $x$ are termed as fast state variables. Detailed power system models and corresponding variables are given in Appendix.

The transient stability model and the QSS model are regarded as two approximations of the long-term stability model in short-term and long-term time scales respectively, and they are believed to offer a good compromise between accuracy and efficiency. In transient stability model, slow variables are considered as constants. While in the QSS model, the dynamic behavior of fast variables are considered as instantaneously fast and thus replaced by its equilibrium equations in the long-term time scale. If we represent the long-term stability model and the QSS model in $\tau$ time scale, where $\tau=t\ee$, and we denote $\prime$ as $\frac{d}{d\tau}$, then power system models can be represented as shown in Table \ref{table1}.

\begin{center}
\begin{table}[h]
\centering
\caption{The mathematical description of models in power system}\label{table1}
\begin{tabular}{|c|c|}
\hline
the long-term stability model&
${z}_{c}^\prime={h}_c({z_c,z_d,x,y})$\\
&${z}_d(k+1)={h}_d({z_c,z_d(k),x,y})$\\
&$\ee{x}^\prime={f}({z_c,z_d,x,y})$\\
&${0}={g}({z_c,z_d,x,y})$\\
\hline
the transient stability model&$\dot{{x}}={f}({z_c,z_d,x,y})$\\
short-term:0-30s&${0}={g}({z_c,z_d,x,y})$\\
\hline
the QSS model&${z}_{c}^\prime={h}_c({z_c,z_d,x,y})$\\
long-term:30s-a few minutes&${z}_d(k+1)={h}_d({z_c,z_d(k),x,y})$\\
&${0}={f}({z_c,z_d,x,y})$\\
&${0}={g}({z_c,z_d,x,y})$\\
\hline
\end{tabular}
\end{table}
\end{center}

The QSS model may fail to capture dynamics of the long-term stability model, thus provide incorrect approximations of the long-term stability model leading to incorrect stability assessment.
\section{Mathematical Preliminaries}\label{mathprelim}
In this section, some relevant stability concepts from nonlinear system theories are briefly reviewed. Knowledge of stability region is required in analyzing the QSS model for long-term stability analysis.

\subsection{Stability of Equilibrium Point and Stability Region}
We consider the following autonomous nonlinear dynamical system:
\begin{equation}\label{general autonomous}
\dot{x}=f(x), \quad x\in\Re^n
\end{equation}
where $f:\Re^n\rightarrow\Re^n$ satisfies a sufficient condition for the existence and uniqueness of a solution. The solution of (\ref{general autonomous}) starting at initial state $x$ at time $t=0$ is called the system trajectory and is denoted as $\phi(t,x)$. $\bar{x}\in\Re^n$ is said to be an equilibrium point of (\ref{general autonomous}) if $f(\bar{x})=0$. The definition of asymptotic stability is given as below \cite{Chiang:book}:

\noindent\textit{\textbf{Definition 1: Asymptotic Stability}}

An equilibrium point $\bar{x}\in\Re^n$ of (\ref{general autonomous}) is said to be asymptotically stable if, for each open neighborhood $U$ of $\bar{x}\in\Re^n$, the followings are true: (i) $\phi(t,x)\in{U}$ for all $t>0$; (ii) $\lim_{t \to \infty}\parallel\phi(t,x)-\bar{x}\parallel=0$.

Without confusion, we use stable equilibrium point (SEP) instead of asymptotically stable equilibrium point in this paper. An equilibrium point is \textit{hyperbolic} if the corresponding Jacobian matrix has no eigenvalues with zero real parts. And a hyperbolic equilibrium point $\bar{x}$ is a \textit{type-k equilibrium point} if there exist $k$ eigenvalues of $D_xf(\bar{x})$ with positive real parts. The stability region of a SEP $x_s$ is the set of all points $x$ such that $\lim_{t \to \infty}\phi(t,x)\rightarrow{x_s}$. In other words, the \textit{stability region} is defined as:

\begin{equation}
A(x_s):=\{x\in\Re^n:\lim_{t \to \infty}\phi(t,x)=x_s\}\nonumber
\end{equation}
From a topological point of view, the stability region is an open invariant and connected set. Every trajectory in a stability region lies entirely in the stability region and the dimension of the stability region is $n$.

\noindent\textit{\textbf{Definition 2: $\omega$-limit Set}}

A point $p$ is said to be the $\omega$-\textit{limit point} of $x$ if, corresponding to each $\ee>0$ and $T>0$, there is a $t>T$ with the property that $||\phi(t,x)-p||<\ee$. Equivalently, there is a sequence ${t_i}$ in $\Re$, $t_i\rightarrow+\infty$, with the property that $p=\lim_{i\rightarrow+\infty}\phi(t_i,x)$. The set of all $\omega$-limit points for $x$ is defined as its $\omega$-\textit{limit set}.

\subsection{Singular Perturbed System}
We next consider the following general singular perturbed model:
\begin{eqnarray}\label{singular perturb}
\Sigma_\ee:\dot{z}&=&f(z,x)\qquad z\in\Re^n\\
\ee\dot{x}&=&g(z,x)\qquad x\in\Re^m\nonumber
\end{eqnarray}
where $\ee$ is a small positive parameter. $z$ is a vector of slow variables while $x$ is a vector of fast variables. Let $\phi_\ee(t,z_0,x_0)$ denotes the trajectory of model (\ref{singular perturb}) starting at $(z_0,x_0)$ and $E$ denotes the set of equilibrium points of it, i.e. $E=\{(z,x)\in\Re^n\times\Re^m: f(z,x)=0, g(z,x)=0\}$. If $(z_s,x_s)$ is a SEP of model (\ref{singular perturb}), then the stability region of $(z_s,x_s)$ is defined as:
\begin{eqnarray}
A_\ee(z_s,x_s):&=&\{(z,x)\in\Re^n\times\Re^m: \phi_\ee(t,z_0,x_0)\rightarrow\nonumber\\
&&(z_s,x_s)\mbox{ as }t \rightarrow\infty\}\nonumber
\end{eqnarray}
The slow model is obtained by setting $\ee=0$ in (\ref{singular perturb}):
\begin{eqnarray}\label{slow}
\Sigma_0:\dot{z}&=&f(z,x)\qquad z\in\Re^n\\
0&=&g(z,x)\qquad x\in\Re^m\nonumber
\end{eqnarray}
The algebraic equation $0=g(z,x)$ constraints the slow dynamics to the following set which is termed as \textit{constraint manifold}:
\begin{equation}
\Gamma:=\{(z,x)\in\Re^n\times\Re^m:g(z,x)=0\}
\end{equation}
The trajectory of model (\ref{slow}) starting at $z_0$ is denoted by $\phi_0(t,z_0,x_0)$ and the stability region is
\begin{equation}
A_0(z_s,x_s):=\{(z,x)\in\Gamma: \phi_0(t,z_0,x_0)\rightarrow(z_s,x_s)\mbox{ as }t\rightarrow\infty\}\nonumber
\end{equation}

The \textit{singular points} of system (\ref{slow}) or \textit{singularity $S$} is defined as:
\begin{equation}
S:=\{(z,x)\in\Gamma:\mbox{ det}(D_xg)(z,x)=0\}
\end{equation}

Singular points can drastically influence the trajectories of the differential-algebraic equation (DAE) system. Typically, the singular set $S$ is a stratified set of maximal dimension $n-1$ embedded in $\Gamma$ and $\Gamma$ is separated by $S$ into open regions \cite{Veukatasubramanian:article}\cite{Alberto:article}.

\noindent \textit{\textbf{Definition 3: Type of Constraint Manifold}}

The connected set $\Gamma_i\subset\Gamma$ is a \textit{type-k component of $\Gamma$} if the matrix $D_xg$, evaluated at every point of $\Gamma_i$, has $k$ eigenvalues that have positive real parts. If all the eigenvalues of $D_xg$ calculated at points of $\Gamma_i$ have a negative real part, then we call $\Gamma_i$ a \textit{stable component of $\Gamma$}; otherwise, it's an \textit{unstable component of $\Gamma$}.

We next define the fast model associated with the singularly perturbed model, i.e. boundary layer model. Define the fast time scale $\sigma=t/\ee$. In this time scale, model (\ref{singular perturb}) takes the form:
\begin{eqnarray}\label{singular perturb in sigma}
\Pi_\ee:\frac{dz}{d\sigma}&=&{\ee}f(z,x)\qquad z\in\Re^n\\
\frac{dx}{d\sigma}&=&g(z,x)\qquad x\in\Re^m\nonumber
\end{eqnarray}
Let $\phi_\ee(\sigma,z_0,x_0)$ denote the trajectory of model (\ref{singular perturb in sigma}) starting at $(z_0,x_0)$.
\begin{equation}\label{BLS}
\Pi_f:\frac{dx}{d\sigma}=g(z,x)
\end{equation}
where $z$ is frozen and treated as a parameter. The constraint manifold $\Gamma$ is a set of equilibriums of models (\ref{BLS}). For each fixed $z$, a fast dynamical model (\ref{BLS}) is defined.

\noindent\textit{\textbf{Definition 4: Uniformly Asymptotically Stable}}

Assuming $(z,x)\notin S$, and $x=j(z)$ is an isolated root of equation:
\begin{equation}\label{BLSsep}
0=g(z,x)
\end{equation}
then $x=j(z)$ is an equilibrium point of system (\ref{BLS}), if $x=j(z)$ is a SEP of system (\ref{BLS}) for all $z\in Z$, then $j(z)$ is \textit{uniformly asymptotically stable} with respect to $z\in Z$.

The next Theorem ensures that, that solutions of the singular perturbed model (\ref{singular perturb}) can be, at least for sufficiently small $\ee$, approximated by solutions of the slow model (\ref{slow}).


\noindent\textbf{Theorem 1 (Tikhonov's Result on Finite Interval)}\cite{Khalil:book}\cite{Lobry:article2}:

Consider the singular perturbation problem (\ref{singular perturb}) and let $x=j(z)$ be an isolated root of system (\ref{BLSsep}). Assume that there exist positive constants $t_1>t_0$, $r$ and $\ee_0$, and a compact domain $Z\subset\Re^n$ such that the following conditions are satisfied for all $t_0\leq t \leq t_1$, $z\in Z$, $||x-j(z)||\leq r$, $0<\ee \leq \ee_0$

(a). The functions $f(z,x)$, $g(z,x)$ and $j(z)$ are continuous;

(b). The slow model (\ref{slow}) has a unique solution $z_0(t)$ with initial condition $z(t_0)=z_0$, defined on $[t_0,t_1]$ and $z_0(t)\in Z$ for all $t\in[t_0,t_1]$;

(c). The fast model (\ref{BLS}) has the uniqueness of the solutions with prescribed initial conditions. Let $\tilde{x}(\sigma)$ be the solution of system:
\begin{equation}\label{initialBLS}
\frac{dx}{d\sigma}=g(z_0,x),\qquad x(\sigma_0)=x_0
\end{equation}

(d). The equilibrium point $x=j(z)$ of fast model is uniformly asymptotically stable in $z\in Z$;

(e). The initial condition $x_0$ belongs to the stability region $A(j(z_0))$ of system (\ref{initialBLS}).

Then for every $\delta>0$, there exists a positive constant $\ee^\star$ such that for all $0<\ee<\ee^\star$, every solution $(z(t),x(t))$ of the singular perturbation model (\ref{singular perturb}) exists at least on $[t_0,t_1]$, and satisfies
\begin{eqnarray}
&&||z(t)-z_0(t)||\leq \delta\nonumber\\
&&||x(t)-\tilde{x}(\frac{t-t_0}{\ee})-j(z_0(t))+j(z_0)||\leq \delta\nonumber
\end{eqnarray}
for all $t_0\leq t \leq t_1$.

Note that $Z\subset\Re^n$ is required to be a compact set, thus the solution of slow model stays inside a compact set to avoid non essential technicalities with the maximal interval of definition of a solution\cite{Lobry:article2}.

Theorem 1 can be extended to the infinite-time interval under some additional conditions which ensure stability of the solutions of the singular perturbation problem (\ref{singular perturb}) \cite{Lobry:article2}.

\noindent\textbf{Theorem 2 (Tikhonov's Result on Infinite Interval)}\cite{Khalil:book}\cite{Lobry:article2}:

Consider the singular perturbation problem (\ref{singular perturb}) and let $x=j(z)$ be an isolated root of system (\ref{BLSsep}). Assume that there exist positive constants $r$ and $\ee_0$, and a compact domain $Z\subset\Re^n$ such that the following conditions are satisfied for all $t_0\leq t \leq +\infty$, $z\in Z$, $||x-j(z)||\leq r$, $0<\ee \leq \ee_0$

(a). The functions $f(z,x)$, $g(z,x)$ and $j(z)$ are continuous;

(b). The solution $z_0(t)$ of the slow model starting from $z(t_0)=z_0$ exists for all $t_0\leq t\leq +\infty$, and the $\omega$-limit set of the slow model (\ref{slow}) is a SEP denoted as $(z_s,x_s)$;

(c). The fast model (\ref{BLS}) has the uniqueness of the solutions with prescribed initial conditions. Let $\tilde{x}(\sigma)$ be the solution of system (\ref{initialBLS});

(d). The equilibrium point $x=j(z)$ of fast model is uniformly asymptotically stable in $z\in Z$;

(e). The initial condition $x_0$ belongs to the stability region $A(j(z_0))$ of system (\ref{initialBLS}).

Then for every $\delta>0$, there exists a positive constant $\ee^\star$ such that for all $0<\ee<\ee^\star$, every solution $(z(t),x(t))$ of the singular perturbation model (\ref{singular perturb}) exists for all $t\geq t_0$, and satisfies
\begin{eqnarray}
&&||z(t)-z_0(t)||\leq \delta\nonumber\\
&&||x(t)-\tilde{x}(\frac{t-t_0}{\ee})-j(z_0(t))+j(z_0)||\leq \delta\nonumber
\end{eqnarray}
for all $t\geq t_0$.

Assume the solution of singular perturbation problem (\ref{singular perturb}) $(z(t,\ee),x(t,\ee))$ is unique, then we have \cite{Lobry:article2}\cite{Lobry:article}:
\begin{eqnarray}
&&\lim_{\ee \to 0\mbox{ } t \to +\infty} z(t,\ee)=\lim_{t \to +\infty} z_0(t)=z_s\\
&&\lim_ {\ee \to 0\mbox{ } t \to +\infty} x(t,\ee)=\lim_{t \to +\infty} j(z_0(t))=x_s\nonumber
\end{eqnarray}

\section{Analytical Studies of QSS model}\label{analyticalstudies}

The long-term stability model of power system can be represented as:
\begin{eqnarray}\label{complete}
{z}_{c}^\prime&=&{h}_c({z_c,z_d,x,y}),\hspace{0.55in}{z_c(\tau_0)=z_{c0}}\\
z_d(k)&=&h_d(z_c,z_d(k-1),x,y),\quad z_d(\tau_0)=z_{d}(0)\nonumber\\
\ee{x}^\prime&=&{f}({z_c,z_d,x,y}),\hspace{0.65in}{x(\tau_0)=x_0^l}\nonumber\\
{0}&=&{g}({z_c,z_d,x,y})\nonumber
\end{eqnarray}

where $\tau=t\ee$. Note that shunt compensation switching and LTC operation are typical discrete events captured by $z_d(k)=h_d(z_c,z_d(k-1),x,y)$ and $z_d$ is shunt susceptance and the transformer ratio, respectively. Transitions of $z_d$ depend on system variables, thus $z_d$ change values from $z_d(k-1)$ to $z_d(k)$ at distinct times $\tau_k$ where $k=1,2,3,...N$, otherwise, these variables remain constants.

Consider the long-term stability model (\ref{complete}), it can be regarded as two decoupled systems (\ref{couple1}) and (\ref{couple2}) shown as below when $z_d$ jump from $z_d(k-1)$ to $z_d(k)$:
\begin{equation}
z_d(k)=h_d(z_c,z_d(k-1),x,y),\qquad z_d(\tau_0)=z_{d}(k-1) \label{couple1}
\end{equation}
and
\begin{eqnarray}\label{couple2}
{z}_{c}^\prime&=&{h}_c({z_c,z_d(k),x,y}),\hspace{0.34in}{z_c(\tau_0)={z}_{ck}}\\
\ee{x}^\prime&=&{f}({z_c,z_d(k),x,y}),\qquad\quad{x(\tau_0)=x_k^l}\nonumber\\
{0}&=&{g}({z_c,z_d(k),x,y})\nonumber
\end{eqnarray}
discrete variables $z_d$ are updated first and then system (\ref{couple2}) works with fixed parameters $z_d$.

Similarly, when $z_d$ jump from $z_d(k-1)$ to $z_d(k)$, the QSS model
\begin{eqnarray}\label{QSS}
{z}_{c}^\prime&=&{h}_c({z_c,z_d,x,y}),\hspace{0.55in}{z_c(\tau_0)=z_{c0}}\\
z_d(k)&=&h_d(z_c,z_d(k-1),x,y),\quad z_d(\tau_0)=z_{d}(0)\nonumber\\
{0}&=&{f}({z_c,z_d,x,y})\nonumber\\
{0}&=&{g}({z_c,z_d,x,y})\nonumber
\end{eqnarray}
can be decoupled as:
\begin{equation}
z_d(k)=h_d(z_c,z_d(k-1),x,y),\qquad z_d(\tau_0)=z_{d}(k-1) \label{coupleqss1}
\end{equation}
and
\begin{eqnarray}\label{coupleqss2}
{z}_{c}^\prime&=&{h}_c({z_c,z_d(k),x,y}),\qquad\quad {z_c(\tau_0)=z_{ck}}\\
{0}&=&{f}({z_c,z_d(k),x,y})\nonumber\\
{0}&=&{g}({z_c,z_d(k),x,y})\nonumber
\end{eqnarray}
\subsection{Models in Nonlinear Framework}
For the study region $U={D_{z_c}}\times{D_{z_d}}\times{D_{x}}\times{D_{y}}$, where $D_{z_c}\subseteq\Re^p$, $D_{z_d}\subseteq\Re^q$, $D_x\subseteq\Re^m$, $D_y\subseteq\Re^n$, both the long-term stability model and the QSS model have the same set of equilibrium points $E=\{(z_c,z_d,x,y)\in{U}:z_d(k)=z_d(k-1),{h}_c({z_{c},z_{d},x,y})=0, {f}({z_c,z_d,x,y})=0,{g}({z_c,z_d,x,y})=0\}$. Assuming $(z_{cls},z_{dls},x_{ls},y_{ls})\in{E}$ is a long-term SEP of both the long-term stability model (\ref{complete}) and the QSS model (\ref{QSS}) starting from $(z_{c0},z_{d}(0),x_0^l,y_0^l)$ and $(z_{c0},z_{d}(0),x_0^q,y_0^q)$ respectively, and $\phi_l(\tau,z_{c0},z_{d}(0),x_0^l,y_0^l)$ denotes trajectory of the long-term stability model (\ref{complete}) and $\phi_q(\tau,z_{c0},z_{d}(0),x_0^q,y_0^q)$ denotes trajectory of the QSS model (\ref{QSS}). Then, the stability region of the long-term stability model (\ref{complete}) is:
\begin{eqnarray}
&&A_l(z_{cls},z_{dls},x_{ls},y_{ls}):=\{(z_c,z_d,x,y)\in{U}:\phi_l(\tau,z_{c0},\nonumber\\
&&z_{d}(0),x_0^l,y_0^l)\rightarrow(z_{cls},z_{dls},x_{ls},y_{ls})\mbox{ as $\tau$}\rightarrow+\infty\}
\end{eqnarray}

The stability region of the QSS model (\ref{QSS}) is
\begin{eqnarray}
&&A_q(z_{cls},z_{dls},x_{ls},y_{ls}):=\{(z_c,z_d,x,y)\in{\Gamma}:\phi_q(\tau,z_{c0},\nonumber\\
&&z_{d}(0),x_0^q,y_0^q)\rightarrow(z_{cls},z_{dls},x_{ls},y_{ls})\mbox{ as $\tau$}\rightarrow+\infty\}
\end{eqnarray}

The singular points of constraint manifold $\Gamma$ are:
\begin{equation}
S:=\{(z_c,z_d,x,y)\in\Gamma:\mbox{det}\left[\begin{array}{cc}D_xf&D_yf\\ D_xg&D_yg\end{array}\right]=0\}
\end{equation}
And type-$k$ component of $\Gamma$ where $0\leq k\leq m+n$ is defined as:
\begin{eqnarray}
&&\Gamma_k=\{(z_c,z_d,x,y)\in\Gamma: \mbox{there are k eigenvalues of}\nonumber\\
&&\left[\begin{array}{cc}D_xf&D_yf\\ D_xg&D_yg\end{array}\right] \mbox{ satisfy Re}(\lambda)>0\}
\end{eqnarray}

When $z_c\in{D_{z_c}}$ and $z_d\in{D_{z_d}}$, for each fixed $z_c$ and $z_d(k)$, given a point $(z_c,z_d(k),x,y)$ on $\Gamma$, the corresponding transient stability model is defined as:
\begin{eqnarray}\label{transient}
\dot{x}&=&{f}({z_c,z_d(k),x,y})\\
{0}&=&{g}({z_c,z_d(k),x,y})\nonumber
\end{eqnarray}
If $(z_c,z_d(k),x,y)\not\in S$, then
$(z_c,z_d(k),x_{ts},y_{ts})$ is an equilibrium point of (\ref{transient}), where
\begin{equation}
\left(\begin{array}{cc}x_{ts}\\y_{ts}\end{array}\right)=\left(\begin{array}{cc}l_1(z_c,z_d(k))\\l_2(z_c,z_d(k))\end{array}\right)=l(z_c,z_d(k))\nonumber
\end{equation}
If $(z_c,z_d(k),x_{ts},y_{ts})$ is a SEP of (\ref{transient}), then the stability region of $(z_c,z_d(k),x_{ts},y_{ts})$ is represented as:
\begin{eqnarray}\label{transientsep}
&&A_t(z_c,z_d(k),x_{ts},y_{ts}):=\{(x,y)\in{D_x}\times{D_y},z_c=z_c,\nonumber\\
&&z_d=z_d(k):\phi_t(t,z_c,z_d(k),x,y)\rightarrow(z_c,z_d(k),x_{ts},y_{ts})\nonumber\\
&&\mbox{as t}\rightarrow+\infty\}\nonumber
\end{eqnarray}
where $\phi_t(t,z_c,z_d(k),x,y)$ denotes the trajectory of the transient stability model (\ref{transient}).

Assuming that $D_y g$ is nonsingular, then transient stability model (\ref{transient}) can be linearized near the equilibrium point as:
\begin{equation}
\dot{x}=(D_x f-D_y f D_y g^{-1}D_x g)x
\end{equation}
and we can define a subset of the stable component of constraint manifold $\Gamma_s\subset\Gamma_0$:

\begin{eqnarray}\label{gammas}
&&\Gamma_s=\{(z_c,z_d,x,y)\in\Gamma: \mbox{all eigenvalues $\lambda$ of }\nonumber\\
&&(D_x f\nonumber-D_y f{D_y g}^{-1}D_x g)\mbox{ satisfy Re}(\lambda)<0, \nonumber\\
&&\mbox{ and }D_y g\mbox{ is nonsingular}\}
\end{eqnarray}
such that each point on $\Gamma_s$ is a SEP of the corresponding transient stability model (\ref{transient}) for fixed $z_c$ and $z_d(k)$. A comprehensive theory of stability regions can be found in \cite{Alberto:article}\cite{Chiang:article1988}\cite{Chiang:article1989}.

We divide the task of establishing a theoretical foundation for the QSS model into two steps as \textbf{Case I} and \textbf{Case II}. Firstly, we analyze the trajectory and $\omega$-limit set relations of the long-term stability model (\ref{couple2}) and the QSS model (\ref{coupleqss2}), that is we regard discrete variables $z_d$ as fixed parameters. Next, we move one step further to include discrete dynamics $z_d$ and deduce the relations of the long-term stability model (\ref{complete}) and the QSS model (\ref{QSS}) in terms of trajectory and $\omega$-limit set.

Before proceeding, we need some important assumptions:

\noindent\textbf{S1.} Neither the long-term stability model nor the QSS model meets singularity points.

\noindent\textbf{S2.} The trajectories of the long-term stability model, the QSS model and transient stability models with specified initial conditions exist and are unique. Additionally, $D_{z_c}$ is compact.

\noindent\textbf{S3.} Equilibrium point of transient stability model is continuous in $z_c$ when $z_d$ are fixed as parameters.

Note that the uniqueness of solutions is generally satisfied in power system models. Besides, since a power system is a real physical system, the domain of each variable is generally compact. As for S3, if S1 is satisfied, we know that equilibrium point of transient stability model $l(z_c,z_d(k))$ is at least locally continuous by Implicit Function Theorem. Moreover, as $z_c$ only varies slowly and subtly, S3 is also generally satisfied. As a result, if S1 is satisfied, we can safely assume that S2 and S3 are satisfied in power system models.

\subsection{Case I: Relations of Trajectory and $\omega$-limit Set}

Assuming the initial point of the long-term stability model is $(z_{ck},z_d(k),x_k^l,y_k^l)$, and the initial point of QSS model is $(z_{ck},z_d(k),x_k^q,y_k^q)$. Then the initial transient stability model can be represented as:
\begin{eqnarray}\label{initialtransientxy}
\dot{x}&=&f(z_{ck},z_d(k),x,y),\qquad x(t_0)=x_k^l\\
0&=&g(z_{ck},z_d(k),x,y)\nonumber
\end{eqnarray}
with the equilibrium point $\left(\begin{array}{cc}x_k^q\\y_k^q\end{array}\right)=\left(\begin{array}{cc}l_1(z_{ck},z_d(k))\\l_2(z_{ck},z_d(k))\end{array}\right)=l(z_{ck},z_d(k))$.
Equivalently, system (\ref{initialtransientxy}) can be represented as
\begin{equation}\label{initialtransient}
\dot{x}=f(z_{ck},z_d(k),x,l_2(z_{ck},z_d(k)))\qquad\quad x(t_0)=x_k^l
\end{equation}
with equilibrium point $x_k^q=l_1(z_{ck},z_d(k))$.

Additionally, denote the solution of QSS model (\ref{coupleqss2}) as $\bar{z}_{ck}(\tau)\in{D_{z_c}}$ and the solution of the initial transient stability model as  $\hat{x}_k(t)$. Besides, denote $D_x^r\subset D_x$ to be a set such that for all $x\in D_x^r$, $||x-l_1(z_{ck},z_d(k))||\leq{r}$, and let $U_r={D_{z_c}}\times{D_{z_d}}\times{D_{x}^r}\times{D_{y}}$.

\noindent\textbf{Theorem 3: (Trajectory Relation)}:

Assuming there exist positive constants $\tau_1>\tau_0$, $r$ and $\ee_0$ such that S1-S3 and the following conditions are satisfied for all $[\tau,z_c,z_d,x,y, \ee]\in[\tau_0,\tau_1]\times{U_r}\times[0,\ee_0]$:

(a). The trajectory $\phi_q(\tau,z_{ck},z_d(k),x_k^q,y_k^q)$ of the QSS model (\ref{coupleqss2}) moves along $\Gamma_s$;

(b). The projection of initial point $(z_{ck},z_d(k),x^l_k,y^l_k)$ of the long-term stability model (\ref{couple2}) to the subspace of $z_{ck}$ and $z_d(k)$ is inside the stability region  of the initial transient stability model (\ref{initialtransient}).

Then for every $\delta>0$ there exists a positive constant $\ee^{\star}$ such that for all $0<\ee<\ee^{\star}$, the solution $(z_{ck}({\tau}), x_k({\tau}))$ of the long-term stability model (\ref{couple2}) exists at least on $[\tau_0, \tau_1]$, and satisfies:
\begin{eqnarray}
&&\|{z}_{ck}({\tau})-\bar{z}_{ck}(\tau)\|\leq{\delta}\\
&&\|x_k({\tau})-l_1(\bar{z}_{ck}(\tau),z_{d}(k))\nonumber\\
&&-\hat{x}_k(\frac{\tau-\tau_0}{\ee})+l_1(z_{ck},z_{d}(k))\|\leq{\delta}\nonumber
\end{eqnarray}
for all $\tau\in[\tau_0,\tau_1]$.

Theorem 3 asserts that if the projection of initial point of the long-term stability model lies inside the stability region of the initial transient stability model, and $\phi_q(\tau,z_{ck},z_d(k),x_k^q,y_k^q)$ moves along $\Gamma_s$, then for sufficiently small $\ee$, trajectory of the long-term stability model (\ref{couple2}) can be approximated by trajectory of the QSS model (\ref{coupleqss2}).

\noindent\textbf{Proof:}
If S1 is satisfied, then $D_yg$ and $\left[\begin{array}{cc}D_xf&D_yf\\ D_xg&D_yg\end{array}\right]$ are nonsingular, according to the Implicit Function Theorem, $x$, $y$ can be solved from:
\begin{eqnarray}
0&=&{f}({z_c,z_d(k),x,y})\\
{0}&=&{g}({z_c,z_d(k),x,y})\nonumber
\end{eqnarray}
with the solution $\left(\begin{array}{cc}x\\y\end{array}\right)=\left(\begin{array}{cc}l_1(z_{c},z_d(k))\\l_2(z_{c},z_d(k))\end{array}\right)=l(z_{c},z_d(k))$, Thus the long-term stability model (\ref{couple2}) becomes:
\begin{eqnarray}\label{approx_tran}
 {{z}}_{c}^\prime&=&{h}_c({z_{c},z_d(k),x,l_2(z_c,z_d(k))})\\
&=&H_c(z_c,z_d(k),x),\qquad{z_c(\tau_0)=z_{ck}}\nonumber\\
\ee{x}^\prime&=&{f}({z_c,z_d(k),x,l_2(z_c,z_d(k))})\nonumber\\
&=&F(z_c,z_d(k),x),\qquad\quad{x(\tau_0)=x_k^l}\nonumber
\end{eqnarray}

Hence, the long-term stability model is transformed into the standard singular perturbation problem considered in Theorem 1, and the QSS model is the corresponding slow model. Next, from the detailed power system models in Appendix, the following fact follows.

\noindent\textit{Fact 1}: These maps $h_c$, $f$ and $g$ that describe slow dynamics, fast dynamics and algebraic constraints respectively are continuous.

With Fact 1 and S3, condition (a) of Theorem 1 is satisfied; with S2, condition (b) and (c) of Theorem 1 are satisfied. Furthermore, if condition (a) of Theorem 3 is satisfied, then $\Gamma_s$ is a subset of stable component of the constraint manifold, thus each point on $\Gamma_s$ is a SEP of the corresponding transient stability model. In other words, if $\phi_q(\tau,z_{ck},z_d(k),x_k^q,y_k^q)$ moves along $\Gamma_s$, then $x=l_1(z_{c},z_d(k))$ is  asymptotically stable uniformly in $z_c$, hence condition (d) of Theorem 1 is satisfied. Note that if $x=l_1(z_{c},z_d(k))$ is asymptotically stable, $x=l_1(z_{c},z_d(k))$ is necessarily to be isolated by definition. Finally, condition (b) of Theorem 3 ensures the satisfaction of condition (e) in Theorem 1. According to Theorem 1, it follows that for every $\delta>0$ there exists a positive constant $\ee^{\star}$ such that for all $0<\ee<\ee^{\star}$, the solution $(z_{ck}({\tau}), x_k({\tau}))$ of system (\ref{approx_tran}) exists at least on $[\tau_0, \tau_1]$, and satisfies:

\begin{eqnarray}
&&\|{z}_{ck}({\tau})-\bar{z}_{ck}(\tau)\|\leq{\delta}, \qquad \tau_0\leq{\tau}\leq{\tau_1}\\
&&\|x_k({\tau})-l_1(\bar{z}_{ck}(\tau),z_{d}(k))-\hat{x}_k(\frac{\tau-\tau_0}{\ee})\nonumber\\
&&+l_1(z_{ck},z_{d}(k))\|\leq{\delta},\qquad\quad\tau_0\leq{\tau}\leq{\tau_1}\nonumber
\end{eqnarray}
This completes the proof of the theorem.

Next we proceed to identify the $\omega$-limit set relation between the long-term stability model (\ref{couple2}) and the QSS model (\ref{coupleqss2}).

\noindent\textbf{Theorem 4: ($\omega$-Limit Set Relation)}:

Assuming there exist positive constants $r$ and $\ee_0$ such that S1-S3 and the following conditions are satisfied for all $[\tau,z_c,z_d,x,y, \ee]\in[\tau_0,+\infty]\times{U_r}\times[0,\ee_0]$:

(a). The trajectory $\phi_q(\tau,z_{ck},z_d(k),x_k^q,y_k^q)$ of the QSS model (\ref{coupleqss2}) moves along $\Gamma_s$;

(b). The projection of initial point $(z_{ck},z_d(k),x^l_k,y^l_k)$ of the long-term stability model (\ref{couple2}) to the subspace of $z_{ck}$ and $z_d(k)$ is inside the stability region  of the initial transient stability model (\ref{initialtransient});

(c). The $\omega$-limit set of the QSS model (\ref{coupleqss2}) starting from $(z_{ck},z_d(k),x_k^q,y_k^q)$ is a SEP $(z_{cks},z_d(k),x_{ks},y_{ks})$.

Then the solution $(z_{ck}({\tau},\ee), x_k({\tau},\ee))$ of the long-term stability model (\ref{couple2}) exists for all $\tau\geq\tau_0$, and satisfies the following limit relations:

\begin{eqnarray}
&&\lim_{\ee \to 0\mbox{ }\tau \to +\infty}z_{ck}(\tau,\ee)={z}_{cks}\\
&& \lim_{\ee \to 0\mbox{ }\tau \to +\infty}x_k(\tau,\ee)=x_{ks}
\end{eqnarray}

Theorem 4 asserts that if all conditions of Theorem 3 are satisfied and the QSS model (\ref{coupleqss2}) converges to a long-term SEP of the QSS model, then for sufficiently small $\ee$, the long-term stability model (\ref{couple2}) will converge to the same point.

\textbf{Proof}:
If S1 is satisfied, then $D_yg$ and $\left[\begin{array}{cc}D_xf&D_yf\\ D_xg&D_yg\end{array}\right]$ are nonsingular. Likewise, we can transform the long-term stability model (\ref{couple2}) to system (\ref{approx_tran}) which is the standard singular perturbation problem considered in Theorem 2.

\begin{figure}[!ht]
\centering
\includegraphics[width=3.5in,keepaspectratio=true,angle=0]{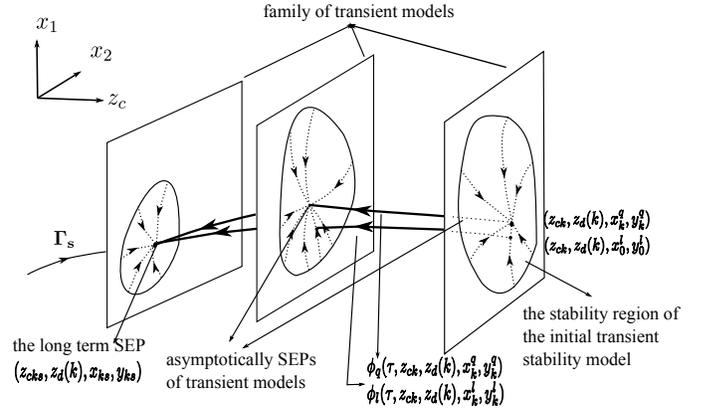}\caption{Illustration of Theorems 3 and 4. $\phi_q(\tau,z_{ck},z_d(k),x_k^q,y_k^q)$ is constrained on $\Gamma_s$ all the time. The projected initial point $(z_{ck},z_d(k),x_k^l,y_k^l)$ of the long-term stability model locates inside $A_t(z_{ck},z_d(k),x_k^q,y_k^q)$, then $\phi_l(\tau,z_{ck},z_d(k),x_k^l,y_k^l)$ always stays close to $\phi_q(\tau,z_{ck},z_d(k),x_k^q,y_k^q)$. Moreover, since $\phi_q(\tau,z_{ck},z_d(k),x_k^q,y_k^q)$ converges to a SEP $(z_{cks},z_{d}(k),x_{ks},y_{ks})$, $\phi_l(\tau,z_{ck},z_d(k),x_k^l,y_k^l)$ also converges to $(z_{cks},z_{d}(k),x_{ks},y_{ks})$.
}\label{QSSctnstable}
\end{figure}

From the proof of Theorem 3, we have that with S2, S3 and Fact 1, condition (a) and (c) of Theorem 2 are satisfied. Besides, condition (a) and (b) of Theorem 4 ensures the satisfaction of condition (d) and (e) in Theorem 2 respectively. Finally, with S2 and condition (c) of Theorem 4, it follows that the solution of the QSS model (\ref{coupleqss2}) exists for all $\tau\geq \tau_0$ and the $\omega$-limit set of the QSS model is a SEP, thus condition (b) of Theorem 2 is satisfied. Therefore all conditions of Theorem 2 are satisfied, it follows that for every $\delta>0$, there exists a positive constant $\ee^\star$ such that for all $0<\ee<\ee^\star$, the solution $({z}_{ck}({\tau}),x_k(\tau))$ of the long-term stability model (\ref{couple2}) exists for all $\tau\geq \tau_0$, and satisfies
\begin{eqnarray}
&&\|{z}_{ck}({\tau})-\bar{z}_{ck}(\tau)\|\leq{\delta}\\
&&\|x_k({\tau})-l_1(\bar{z}_{ck}(\tau),z_{d}(k))-\hat{x}_k(\frac{\tau-\tau_0}{\ee})\nonumber\\
&&+l_1(z_{ck},z_{d}(k))\|\leq{\delta}\nonumber
\end{eqnarray}
for all $\tau\geq \tau_0$. Since the solution of the long-term stability model (\ref{couple2}) $(z_{ck}(\tau,\ee),x_k(\tau,\ee))$is unique, we have
\begin{eqnarray}
&&\lim_{\ee \to 0\mbox{ }\tau \to +\infty}z_{ck}(\tau,\ee)={z}_{cks}\\
&& \lim_{\ee \to 0\mbox{ }\tau \to +\infty}x_k(\tau,\ee)=x_{ks}
\end{eqnarray}
This completes the proof of the theorem. And Fig. \ref{QSSctnstable} gives an illustration of Theorem 3 and Theorem 4.

\subsection{Case II: Relations of Trajectory and $\omega$-limit Set}

Next, we are at the stage to incorporate discrete behaviors of $z_d$ in the long-term stability model and the QSS model, and explore trajectory and $\omega$-limit set relations between them.

Assuming $z_d=z_d(0)$ initially at $\tau_0$, and jump from $z_d(k-1)$ to $z_d(k)$ at time $\tau_k$, where $k=1,2,3...N$. Similarly, the initial point of the long-term stability model is $(z_{c0},z_d(0),x_0^l,y_0^l)$, and the initial point of QSS model is $(z_{c0},z_d(0),x_0^q,y_0^q)$. Then the initial transient stability model can be represented as:
\begin{equation}\label{initialtransient_II}
\dot{x}=f(z_{c0},z_d(0),x,l_2(z_{c0},z_d(0)))\qquad\quad x(t_0)=x_0^l
\end{equation}
with equilibrium point $x_0^q=l_1(z_{c0},z_d(0))$.

Denote the solution of QSS model (\ref{coupleqss2}) as $\bar{z}_{ck}(\tau)\in{D_{z_c}}$, and denote the solution of the initial transient stability model and transient stability models immediately after $z_d$ jump to $z_d(k)$ as $\hat{x}_k(\frac{\tau-\tau_k}{\ee})$ for all $k=0,1,2...N$.

\noindent\textit{\textbf{Definition 5: Consistent Attraction}}

We say that the long-term stability model satisfies the condition of \textit{consistent attraction}, if whenever long-term discrete variables jump from $z_d(k-1)$ to $z_d(k)$, $k=1,2,3...N$, the point on the trajectory of the long-term stability model immediately after $z_d$ jump stays inside the stability region of the corresponding transient stability model.

The following two theorems provide a theoretical foundation for the QSS model in which trajectory and $\omega$-limit set relations of the long-term stability model (\ref{complete}) and the QSS model (\ref{QSS}) are established.

\noindent\textbf{Theorem 5: (Trajectory Relation)}

Assuming there exist positive constants $\tau_1>\tau_0$, $r$ and $\ee_0$ such that S1-S3 and the following conditions are satisfied for all $[\tau,z_c,z_d,x,y, \ee]\in[\tau_0,\tau_1]\times{U_r}\times[0,\ee_0]$:

(a). The trajectory $\phi_q(\tau,z_{c0},z_d(0),x_0^q,y_0^q)$ of the QSS model (\ref{QSS}) moves along $\Gamma_s$;

(b). The projection of initial point $(z_{c0},z_d(0),x^l_0,y^l_0)$ of the long-term stability model (\ref{complete}) to the subspace of $z_{c0}$ and $z_d(0)$ is inside the stability region  of the initial transient stability model (\ref{initialtransient_II}), and the long-term stability model (\ref{complete}) satisfies the condition of consistent attraction.

Then for every $\delta>0$ there exists a positive constant $\ee^{\star}$ such that for all $0<\ee<\ee^{\star}$, every solution $(z_{ck}({\tau}), x_k({\tau}))$ of system (\ref{couple2}) exists at least on $[\tau_k, \tau_{k+1}]$, and satisfies:
\begin{eqnarray}
&&\|{z}_{ck}({\tau})-\bar{z}_{ck}(\tau)\|\leq{\delta}\\
&&\|x_k({\tau})-l_1(\bar{z}_{ck}(\tau),z_{d}(k))\nonumber\\
&&-\hat{x}(\frac{\tau-\tau_k}{\ee})+l_1(\tilde{z}_{ck},z_{d}(k))\|\leq{\delta}\nonumber
\end{eqnarray}
for all $\tau\in[\tau_k,\tau_{k+1}]$, $k\in[0,1,2...N]$.

Theorem 5 asserts that if the trajectory of QSS model moves along $\Gamma_s$, and the projection of each point on trajectory of the long-term stability model always lies inside the stability region of the corresponding transient stability model, then for sufficiently small $\ee$, trajectory of the long-term stability model (\ref{complete}) can be approximated by trajectory of the QSS model (\ref{QSS}).

\noindent\textbf{Proof:}
Conditions of Theorem 5 ensure that conditions of Theorem 3 are satisfied for each fixed $z_d(k)$, $k=0,1,2...N$. Thus we can apply the conclusions of Theorem 3 for each $z_d(k)$. We have that, for every $\delta>0$ there exists a positive constant $\ee_k$ such that for all $0<\ee<\ee_k$, the solution $(z_{ck}({\tau}), x_k({\tau}))$ of system (\ref{couple2}) exists at least on $[\tau_k, \tau_{k+1}]$, and satisfies:
\begin{eqnarray}
&&\|{z}_{ck}({\tau})-\bar{z}_{ck}(\tau)\|\leq{\delta}\\
&&\|x_k({\tau})-l_1(\bar{z}_{ck}(\tau),z_{d}(k))\nonumber\\
&&-\hat{x}(\frac{\tau-\tau_k}{\ee})+l_1(\tilde{z}_{ck},z_{d}(k))\|\leq{\delta}\nonumber
\end{eqnarray}
for all $\tau\in[\tau_k,\tau_{k+1}]$. Let $\ee^\star=\mbox{min}(\ee_0,\ee_1,...\ee_N)$, then for every $\delta>0$ there exists a positive constant $\ee^{\star}$ such that for all $0<\ee<\ee^{\star}$, the solution $(z_{ck}({\tau}), x_k({\tau}))$ of system (\ref{couple2}) exists at least on $[\tau_k, \tau_{k+1}]$, and satisfies:
\begin{eqnarray}
&&\|{z}_{ck}({\tau})-\bar{z}_{ck}(\tau)\|\leq{\delta}\\
&&\|x_k({\tau})-l_1(\bar{z}_{ck}(\tau),z_{d}(k))\nonumber\\
&&-\hat{x}(\frac{\tau-\tau_k}{\ee})+l_1(\tilde{z}_{ck},z_{d}(k))\|\leq{\delta}\nonumber
\end{eqnarray}
for all $\tau\in[\tau_k,\tau_{k+1}]$, where $k\in[0,1,2...N]$.
The proof is complete.

We next show the $\omega$-limit set relation between the long-term stability model (\ref{complete}) and the QSS model (\ref{QSS}).

\noindent\textbf{Theorem 6: ($\omega$-Limit Set Relation)}

Assuming there exist positive constants $r$ and $\ee_0$ such that S1-S3 and the following conditions are satisfied for all $[\tau,z_c,z_d,x,y, \ee]\in[\tau_0,+\infty]\times{U_r}\times[0,\ee_0]$:

(a). The trajectory $\phi_q(\tau,z_{c0},z_d(0),x_0^q,y_0^q)$ of the QSS model (\ref{QSS}) moves along $\Gamma_s$;

(b). The projection of initial point $(z_{c0},z_d(0),x^l_0,y^l_0)$ of the long-term stability model (\ref{complete}) to the subspace of $z_{c0}$ and $z_d(0)$ is inside the stability region  of the initial transient stability model (\ref{initialtransient_II}), and the long-term stability model (\ref{complete}) satisfies the condition of consistent attraction;

(c). The $\omega$-limit set of the QSS model (\ref{QSS}) starting from $(z_{c0},z_d(0),x_0^q,y_0^q)$ is a SEP $(z_{cls},z_{dls},x_{ls},y_{ls})$.

Then the solution $(z_c({\tau},\ee), x({\tau},\ee))$ of the long-term stability model (\ref{complete}) exists for all $\tau\geq\tau_0$, and satisfies the following limit relations:
\begin{eqnarray}
&&\lim_{\ee \to 0\mbox{ }\tau \to +\infty}z_{c}(\tau,\ee)={z}_{cls}\\
&& \lim_{\ee \to 0\mbox{ }\tau \to +\infty}x(\tau,\ee)=x_{ls}
\end{eqnarray}

Theorem 6 asserts that if all conditions of Theorem 5 are satisfied and the QSS model (\ref{QSS}) converges to a long-term SEP of the QSS model, then for sufficiently small $\ee$, the long-term stability model will converge to the same point.

\noindent\textbf{Proof:}
Since $(z_{cNs},x_{Ns})=(z_{cls},x_{ls})$, then according to Theorem 4, we have:
\begin{eqnarray}
\label{k=N_1}
&&\lim_{\ee \to 0\mbox{ } \tau \to +\infty}z_{cN}(\tau,\ee)=z_{cNs}={z}_{cls}\nonumber\\
&& \lim_{\ee \to 0\mbox{ } \tau \to +\infty}x_{N}(\tau,\ee)=x_{Ns}=x_{ls}\nonumber
\end{eqnarray}

Next, since the long-term stability model (\ref{approx_tran}) with each fixed parameter $z_{d}(k)$ has a unique solution for all $k=0,1,2,....N$, the whole long-term stability model (\ref{complete}) with initial condition $(z_{c0},z_d(0),x_0^l,y_0^l)$ will also has a unique solution which is denoted as $(z_{c}(\tau,\ee), x(\tau,\ee))$. Hence we have:
\begin{eqnarray}
\label{k=N}
&&\lim_{\ee \to 0\mbox{ }  \tau \to +\infty}z_c(\tau,\ee)=\lim_{\ee \to 0\mbox{ } \tau \to +\infty}z_{cN}(\tau,\ee)=z_{cls}\nonumber\\
&&\lim_{\ee \to 0\mbox{ }  \tau \to +\infty}x(\tau,\ee)=\lim_{\ee \to 0\mbox{ } \tau \to +\infty}x_{N}(\tau,\ee)=x_{ls}\nonumber
\end{eqnarray}

The proof is complete.

\begin{figure}[!ht]
\centering
\includegraphics[width=3.3in,keepaspectratio=true,angle=0]{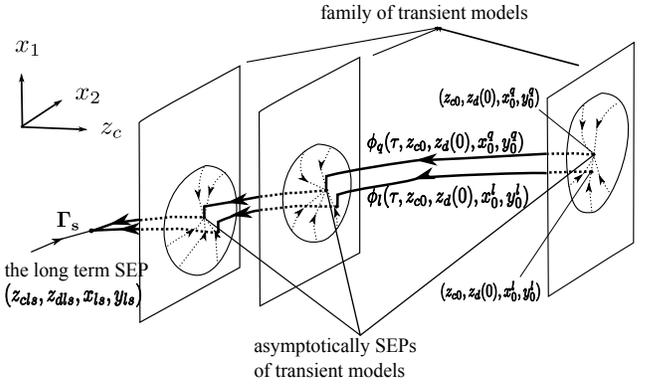}\caption{Illustration of Theorems 5 and 6. $\phi_q(\tau,z_{c0},z_d(0),x_0^q,y_0^q)$ is constrained on $\Gamma_s$ all the time. The projected initial point of the long-term stability model locates inside the stability region of the initial transient stability model, and the long-term stability model satisfies the condition of consistent attraction. Finally both $\phi_q(\tau,z_{c0},z_d(0),x_0^q,y_0^q)$ and $\phi_l(\tau,z_{c0},z_d(0),x_0^l,y_0^l)$ converge to the same point $(z_{cls},z_{dls},x_{ls},y_{ls})$. }\label{QSSstable}
\end{figure}

\begin{figure}[!ht]
\centering
\includegraphics[width=3.5in,keepaspectratio=true,angle=0]{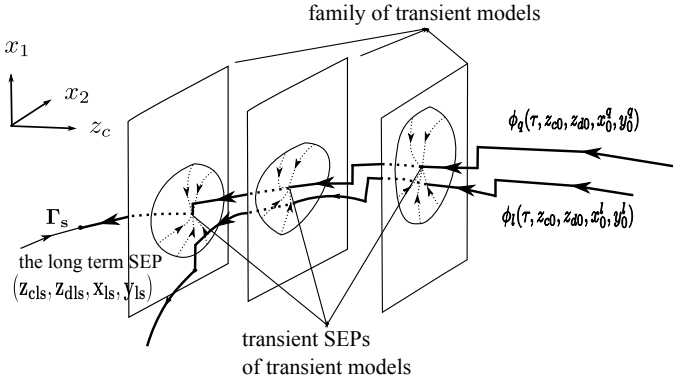}\caption{Illustration of Theorems 5 and 6. When $z_d$ jump to $z_d(k)$, the first point on $\phi_l(\tau,z_{c0},z_d(0),x_0^l,y_0^l)$ after the jump gets outside of the stability region of the corresponding transient stability model, thus the long-term stability model doesn't satisfy the condition of consistent attraction, and $\phi_l(\tau,z_{c0},z_d(0),x_0^l,y_0^l)$ moves far away from the QSS model from then on.}\label{QSSzdunstable}
\end{figure}

Fig. \ref{QSSstable} gives an illustration of Theorem 5 and Theorem 6. Note that the condition of consistent attraction is crucial. For instance, assuming when $z_d$ jump from $z_d(k-1)$ to $z_d(k)$, the first point after the jump $({z}_{ck},z_{d}(k),{x}_k,{y}_k)$ on $\phi_l(\tau,z_{c0},z_d(0),x^l_0,y^l_0)$ locates outside the stability region $A_t({z}_{ck},z_d(k),x^q_k,y^q_k)$ of the transient stability model:
\begin{eqnarray}\label{transient2}
\dot{x}&=&{f}({{z}_c,z_{d}(k),x,y}), \qquad x(t_0)=x^q_k\\
{0}&=&{g}({{z}_c,z_{d}(k),x,y})\nonumber
\end{eqnarray}
then $\phi_l(\tau,z_{c0},z_d(0),x^l_0,y^l_0)$ will move away from $\Gamma_s$ as shown in Fig. \ref{QSSzdunstable}.

\section{Numerical Illustration} \label{numericalstudies}
In this section, two examples will be analyzed using the derived theorems. In the first numerical example, the QSS model provided correct approximations of the long-term stability model since all conditions of Theorem 5 and Theorem 6 are satisfied. In the second numerical example, the QSS model failed to give correct approximations due to the violation of condition (b) in Theorem 5. All simulations were done using PSAT 2.1.6 \cite{Milano:article}.
\subsection{Numerical Example I}
\begin{figure}[!ht]
\centering
\begin{minipage}[t]{0.5\linewidth}
\includegraphics[width=1.8in ,keepaspectratio=true,angle=0]{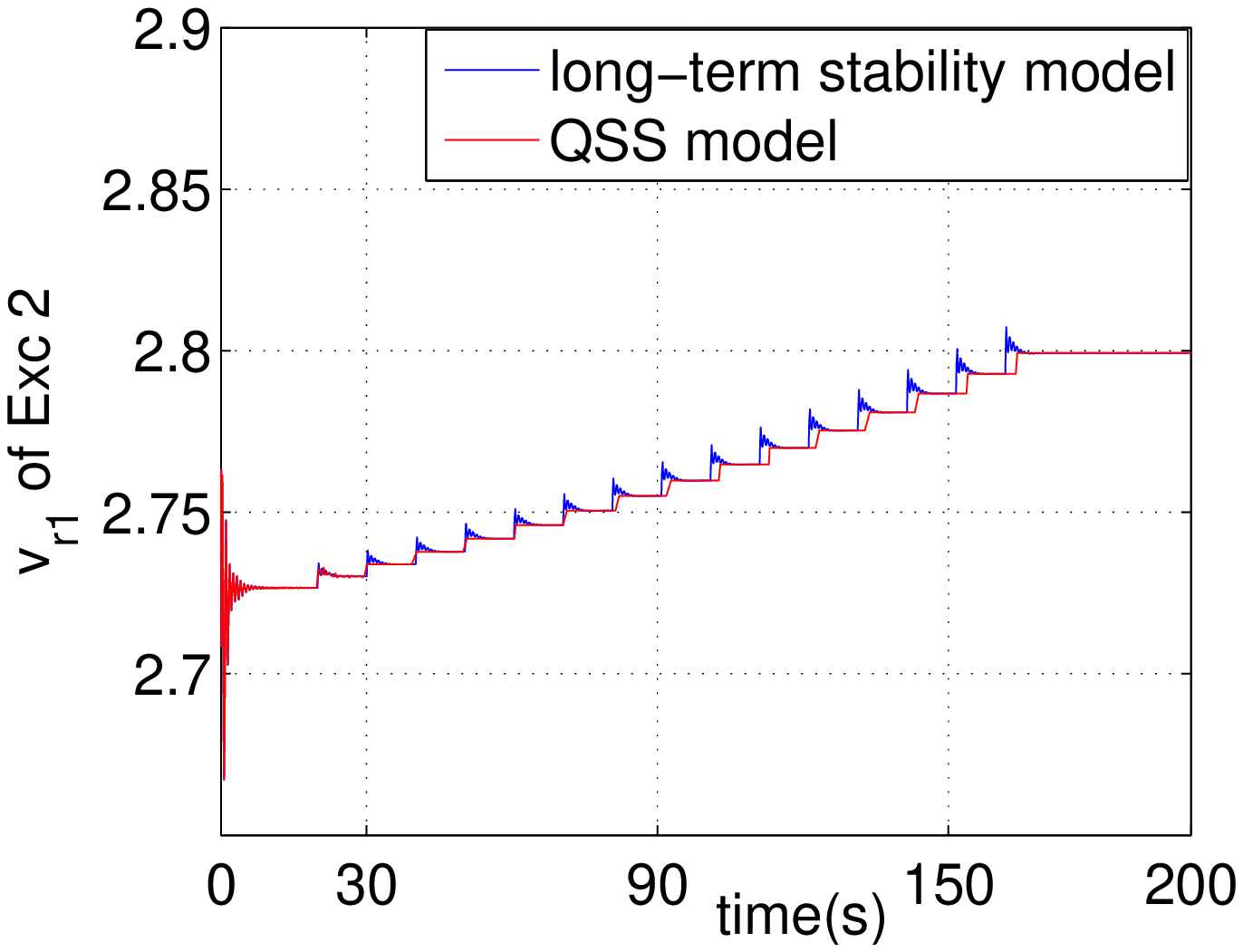}
\end{minipage}%
\begin{minipage}[t]{0.5\linewidth}
\includegraphics[width=1.8in ,keepaspectratio=true,angle=0]{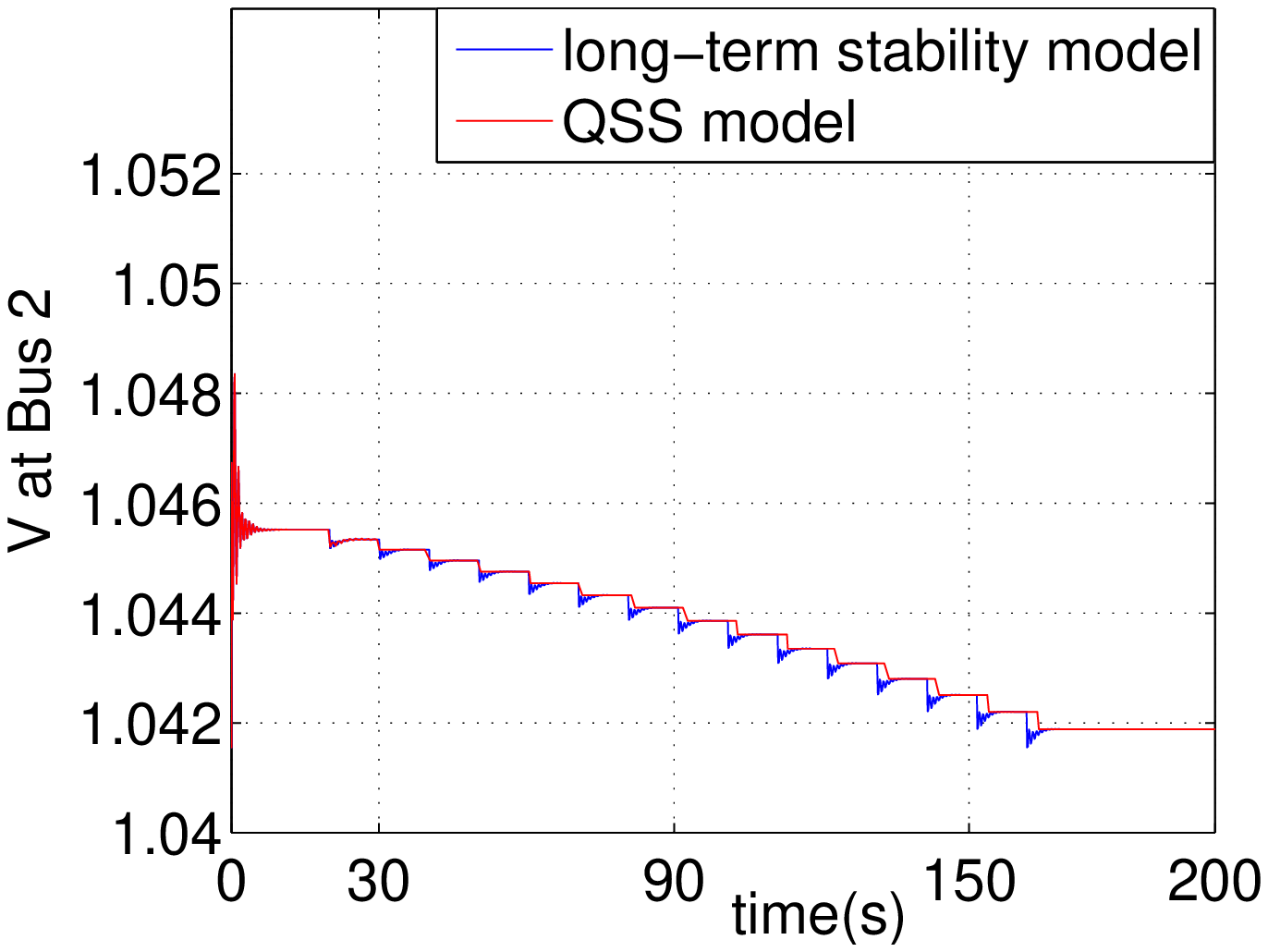}
\end{minipage}
\caption{The trajectory comparisons of the long-term stability model and the QSS model. The trajectory of the long-term stability model followed that of the QSS model until both of them converged to the same long-term SEP.}\label{my14completeqss}
\end{figure}

The first example was a modified IEEE 14-bus systems\cite{testcasearchive} in which QSS model gave correct approximations of the long-term stability model. In this system, an exponential recovery load was included at Bus 5 and two turbine governors at Bus 1 and Bus 2 were added respectively. The assumption S1 that neither the long-term stability model and the QSS model meets singularity points was satisfied. And we can safely assume that S2 was also satisfied. Besides, from the trajectory of the QSS model, we can see that S3 was also satisfied in this case. In addition, the trajectory $\phi_q(\tau,z_{c0},z_d(0),x_0^q,y_0^q)$ of QSS model moved along $\Gamma_s$. As QSS model was implemented 30s after the contingency, fast dynamics settled down at the time such that the projection of the initial point $(z_{c0},z_{d}(0),x_0^l,y_0^l)$ of the long-term stability model lied inside $A_t(z_{c0},z_{d}(0),l_1(z_{c0},z_{d}(0)),l_2(z_{c0},z_{d}(0)))$. And whenever $z_d$ jumped to $z_d(k)$, $k=1,2,...N$, the first point $({z}_{ck},z_{d}(k),{x}_k,{y}_k)$ on  $\phi_l(\tau,z_{c0},z_d(0),x_0^l,y_0^l)$ after the jump stayed inside the stability region of the corresponding transient stability model such that the long-term stability model satisfied the condition of consistent attraction. Since all conditions of Theorem 5 were satisfied, $\phi_l(\tau,z_{c0},z_d(0),x_0^l,y_0^l)$ always stayed close to $\phi_q(\tau,z_{c0},z_d(0),x_0^q,y_0^q)$. Additionally, as the QSS model converged to a long-term SEP, the long-term stability model converged to the same point. Fig. \ref{my14completeqss} shows the trajectory comparisons of the long-term stability model and the QSS model.

\begin{figure}
\centering
\subfloat[]{\label{vr1vf_transient_30}\includegraphics[width=1.8in ,keepaspectratio=true,angle=0]{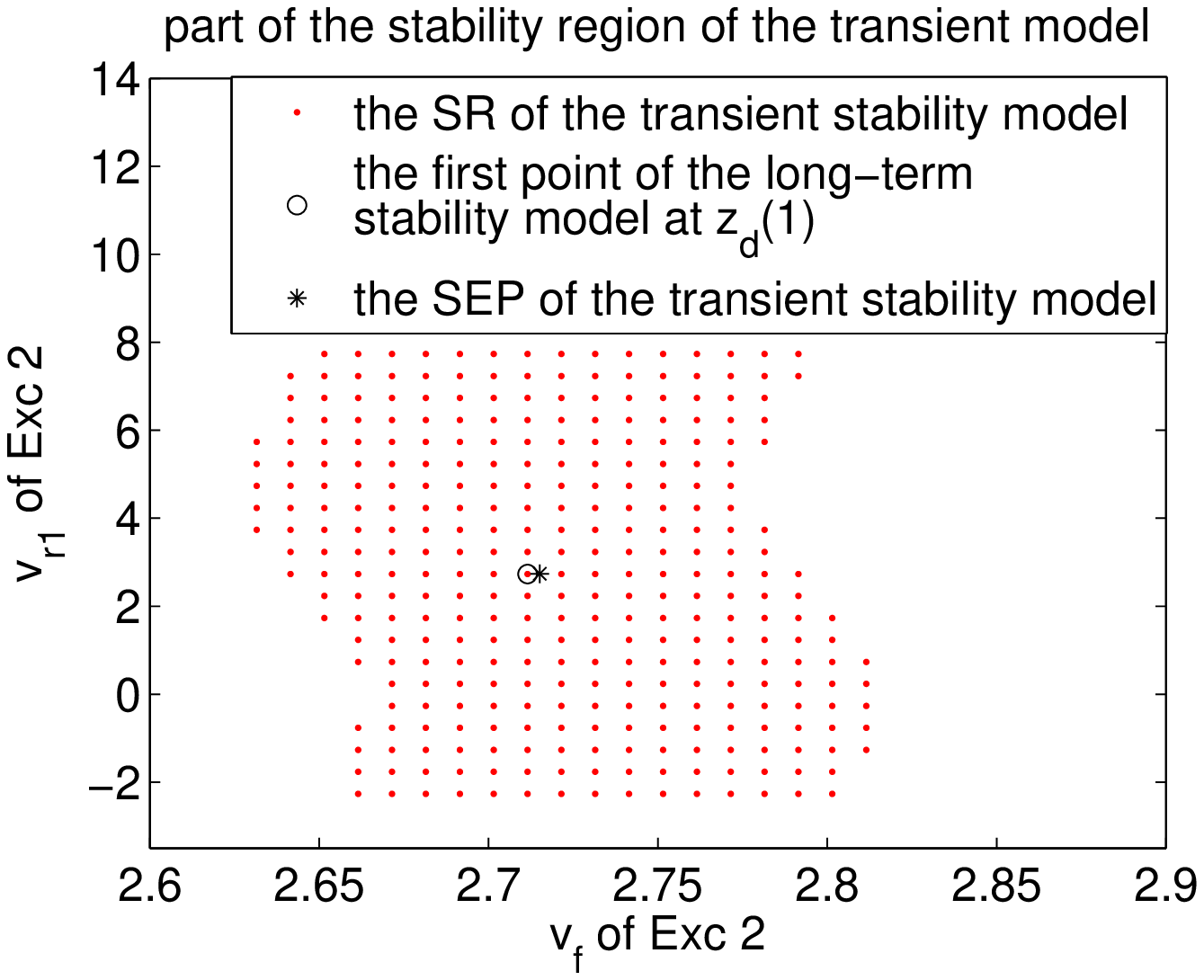}}
\subfloat[]{\label{vr1vf_transient_50}\includegraphics[width=1.8in ,keepaspectratio=true,angle=0]{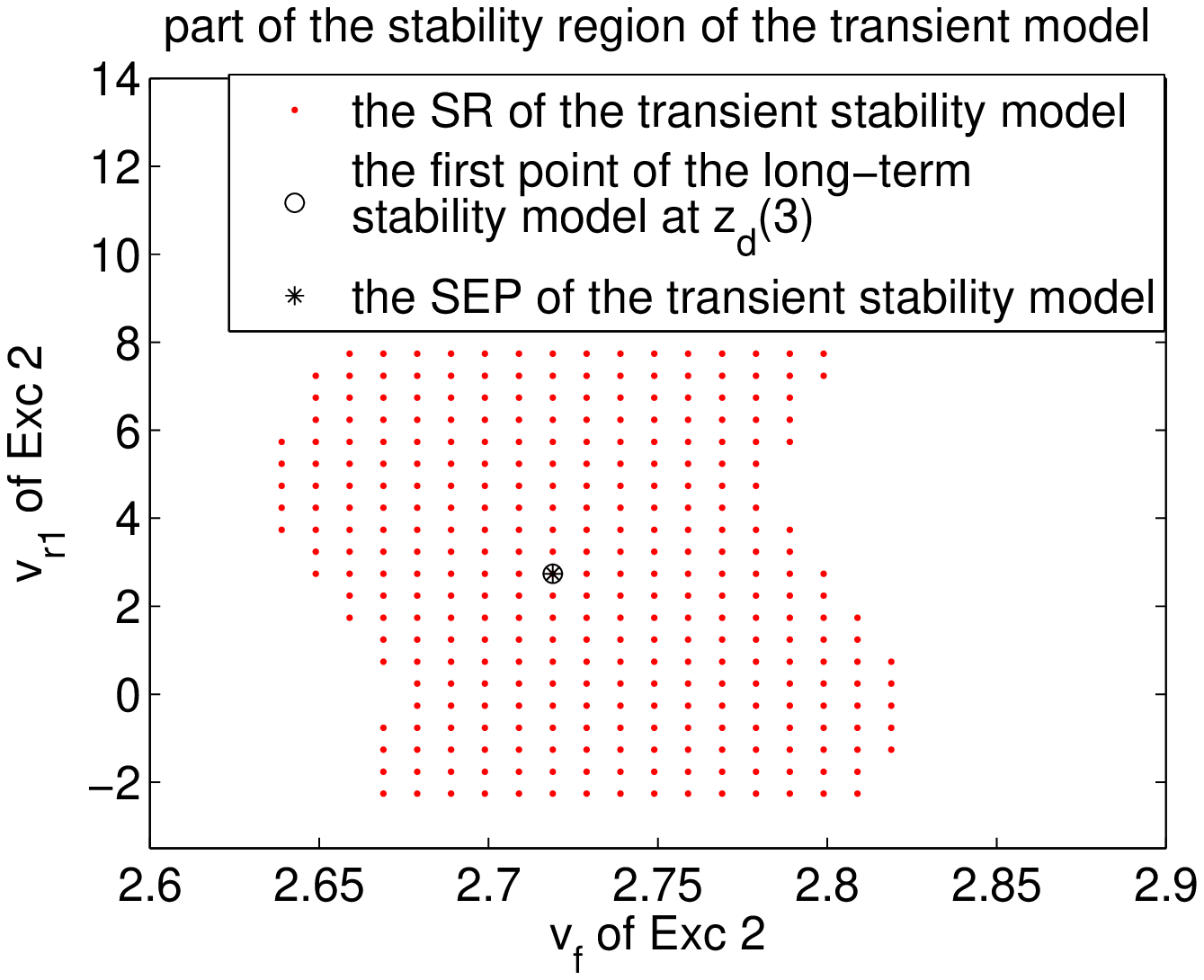}}
\caption{Illustration of Theorems 5 and 6. (a). The stability region of the corresponding transient stability models in the subspace of two fast variables when $z_d$ changed to $z_d(1)$ at 30s; (b). The same as (a), except that $z_d$ changed to $z_d(3)$ at 50s. In both (a) and (b), the first points of the long-term stability model after $z_d$ changed were inside the stability region of the corresponding transient stability models.}
\label{my14_SR}
\end{figure}

\begin{figure}[!ht]
\begin{minipage}[t]{0.5\linewidth}
\includegraphics[width=1.8in ,keepaspectratio=true,angle=0]{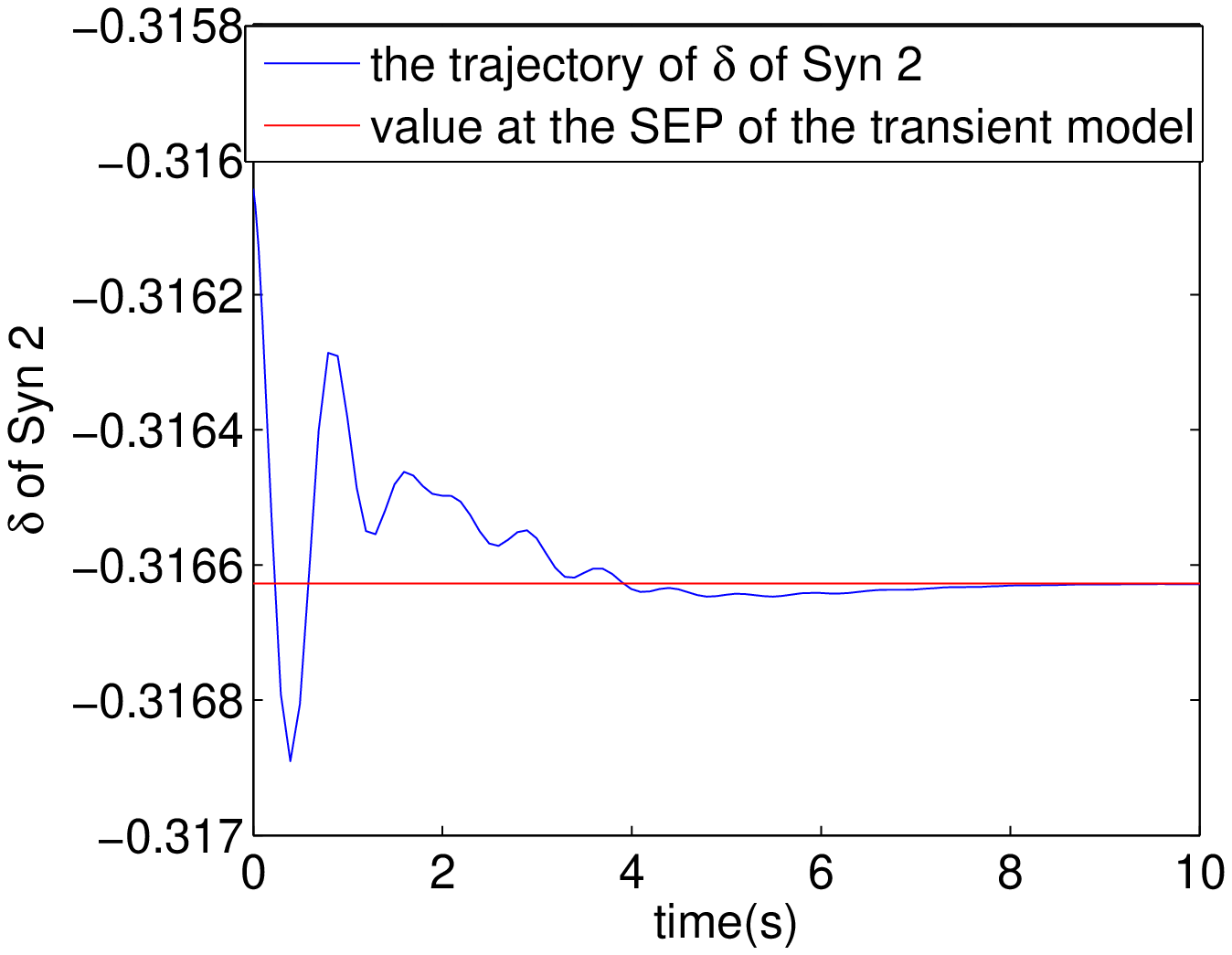}
\end{minipage}%
\begin{minipage}[t]{0.5\linewidth}
\includegraphics[width=1.8in,keepaspectratio=true,angle=0]{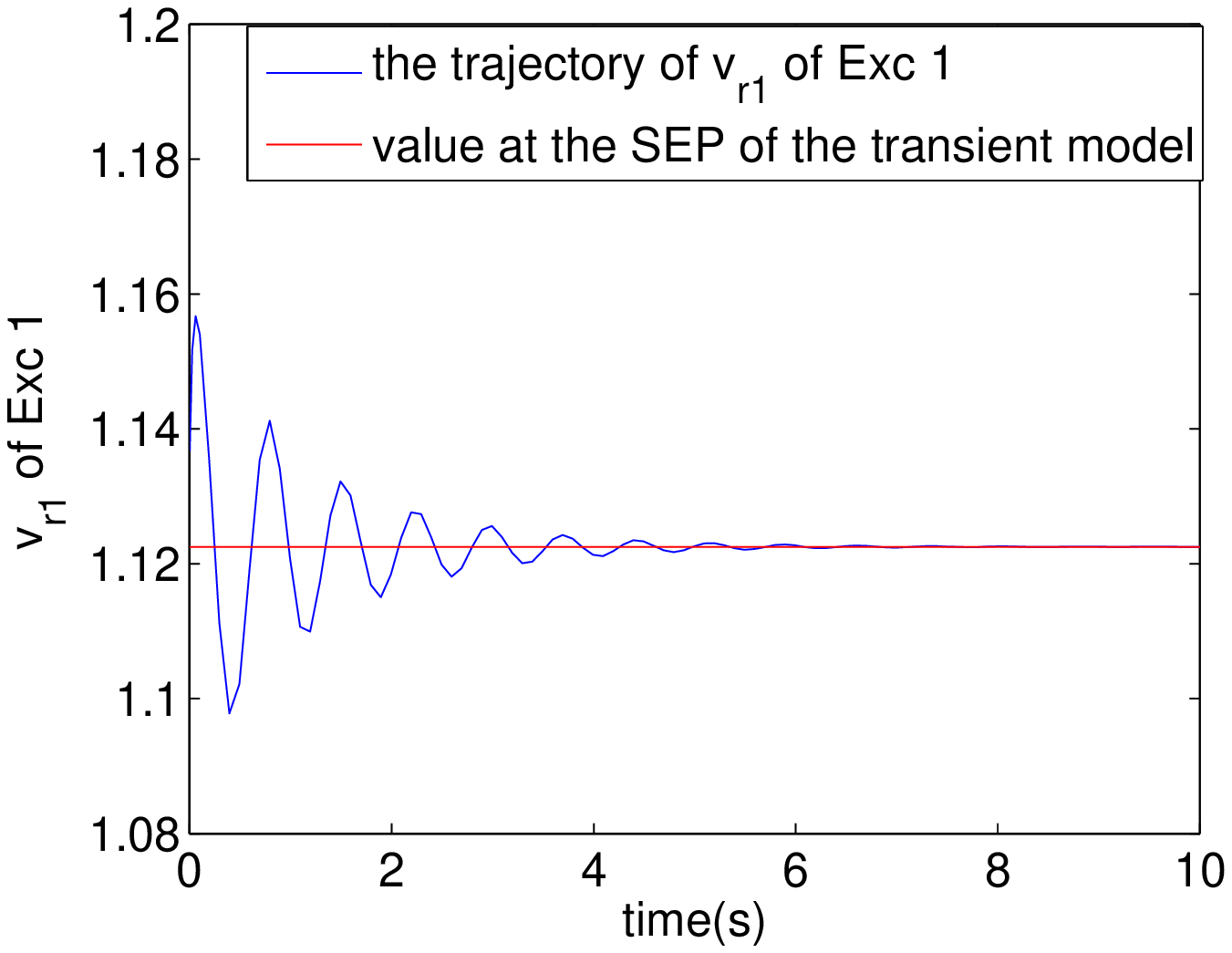}
\end{minipage}
\caption{Trajectories of the transient stability model when $z_d$ jumped to $z_d(1)$ at 30s. The trajectories starting from the first point of the long-term stability model converged to the SEP of the transient stability model which indicated that the first point of the long-term stability model was inside the stability region of the transient stability model.}\label{my14transient0.959}
\end{figure}

To check the condition of consistent attraction, we did the following simulations. When the QSS model was implemented and the ratio of the LTC firstly jumped at 30s, the intersection of the stability region in the subspace of two fast variables is plotted in Fig. \ref{vr1vf_transient_30}, and the first point $({z}_{c1},z_{d}(1),{x}_1,{y}_1)$ on $\phi_l(\tau,z_{c0},z_d(0),x_0^l,y_0^l)$ when $z_d$ jumped to $z_d(1)$ was marked. Additionally, trajectories of two fast variables in the corresponding transient stability model are shown in Fig. \ref{my14transient0.959}. It can be seen that the trajectory starting from $({z}_{c1},z_{d}(1),{x}_1,{y}_1)$ settled down to the SEP of the transient stability model which further confirmed that $({z}_{c1},z_{d}(1),{x}_1,{y}_1)$ did lie inside the stability region $A_t({z}_{c1},z_{d}(1),l_1({z}_{c1},z_{d}(1)),l_2({z}_{c1},z_{d}(1)))$ of the corresponding transient stability model.

Fig. \ref{vr1vf_transient_50} shows stability region of the transient stability model in the subspace of the same fast variables when $z_d$ jumped from $z_d(2)$ to $z_d(3)$. Likewise, this procedure can be done successively to verify that the long-term stability model satisfied the condition of consistent attraction.

\subsection{Numerical Example II}
\begin{figure}[!ht]
\centering
\begin{minipage}[t]{0.5\linewidth}
\includegraphics[width=1.8in ,keepaspectratio=true,angle=0]{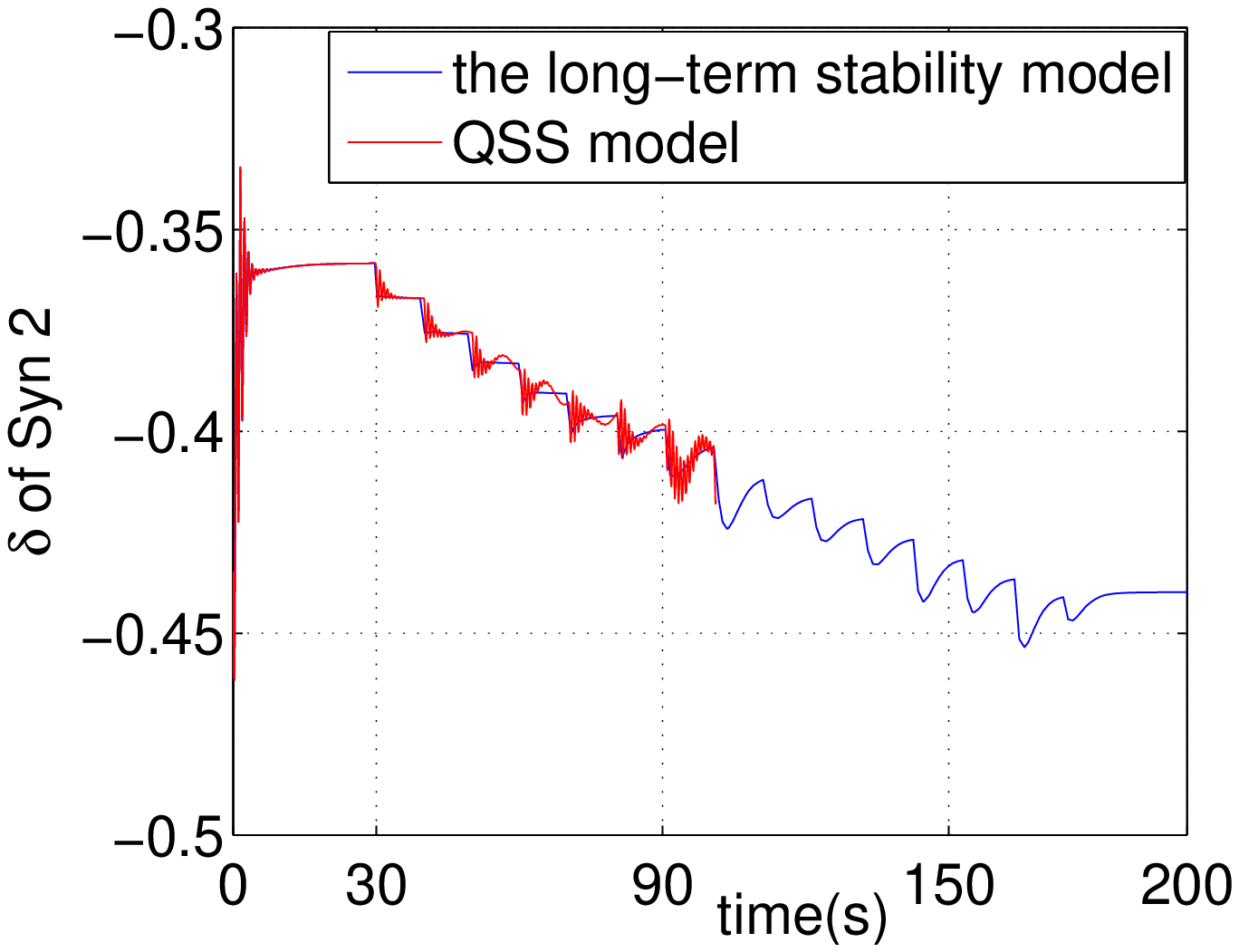}
\end{minipage}%
\begin{minipage}[t]{0.5\linewidth}
\includegraphics[width=1.8in ,keepaspectratio=true,angle=0]{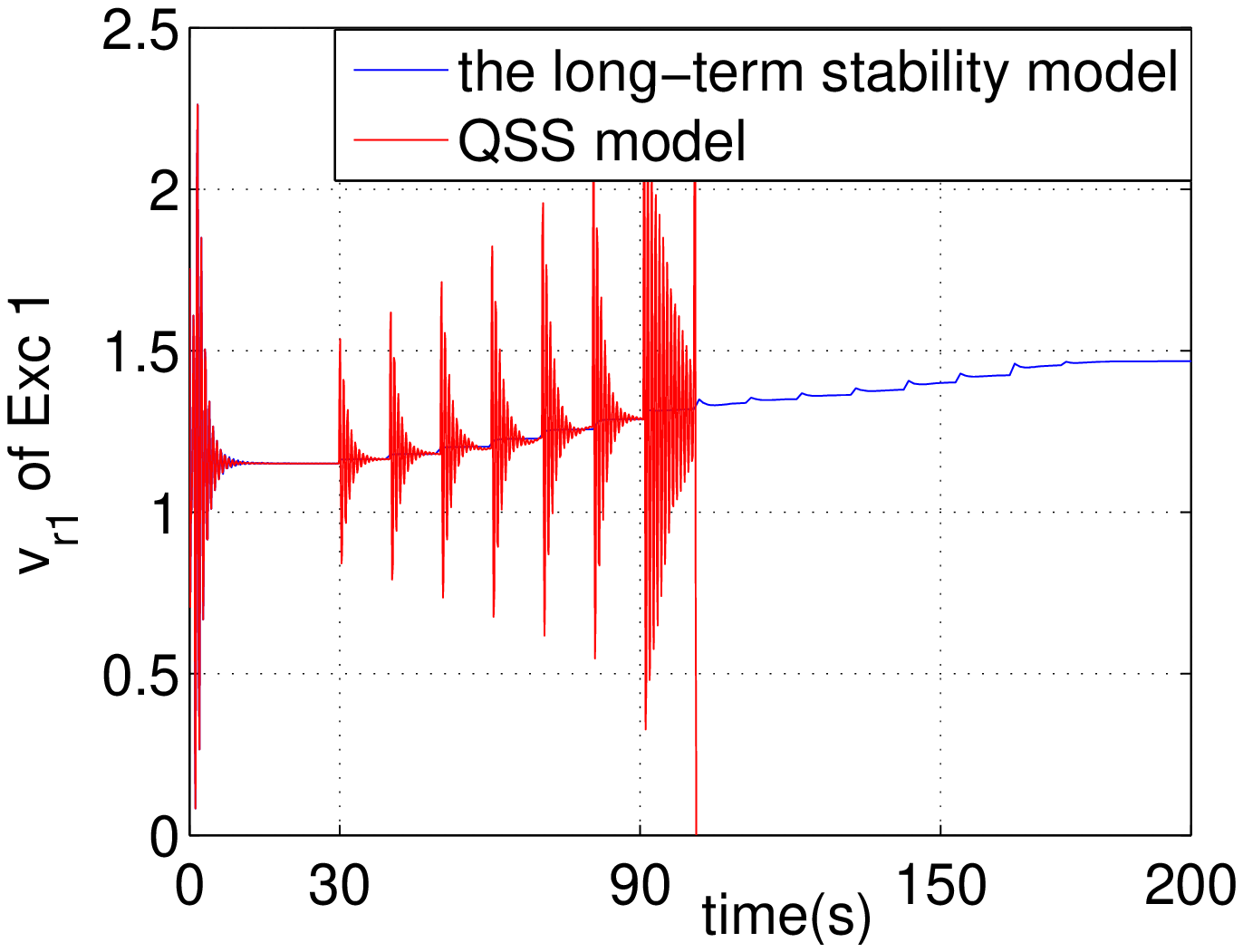}
\end{minipage}
\caption{The trajectory comparisons of the long-term stability model and the QSS model. The QSS model converged to a long-term SEP while the long-term stability model stopped at 101.2155s s due to instability caused by wild oscillation of transient variables.}\label{my14completeqss_try}
\end{figure}

This was also a 14-bus system, while the QSS model did not give correct approximations of the long-term stability model due to the violation of condition (b) in Theorem 5. The trajectory comparisons are shown in Fig. \ref{my14completeqss_try}.

\begin{figure}%
\centering
\subfloat[]{\label{my14try_31}\includegraphics[width=1.8in ,keepaspectratio=true,angle=0]{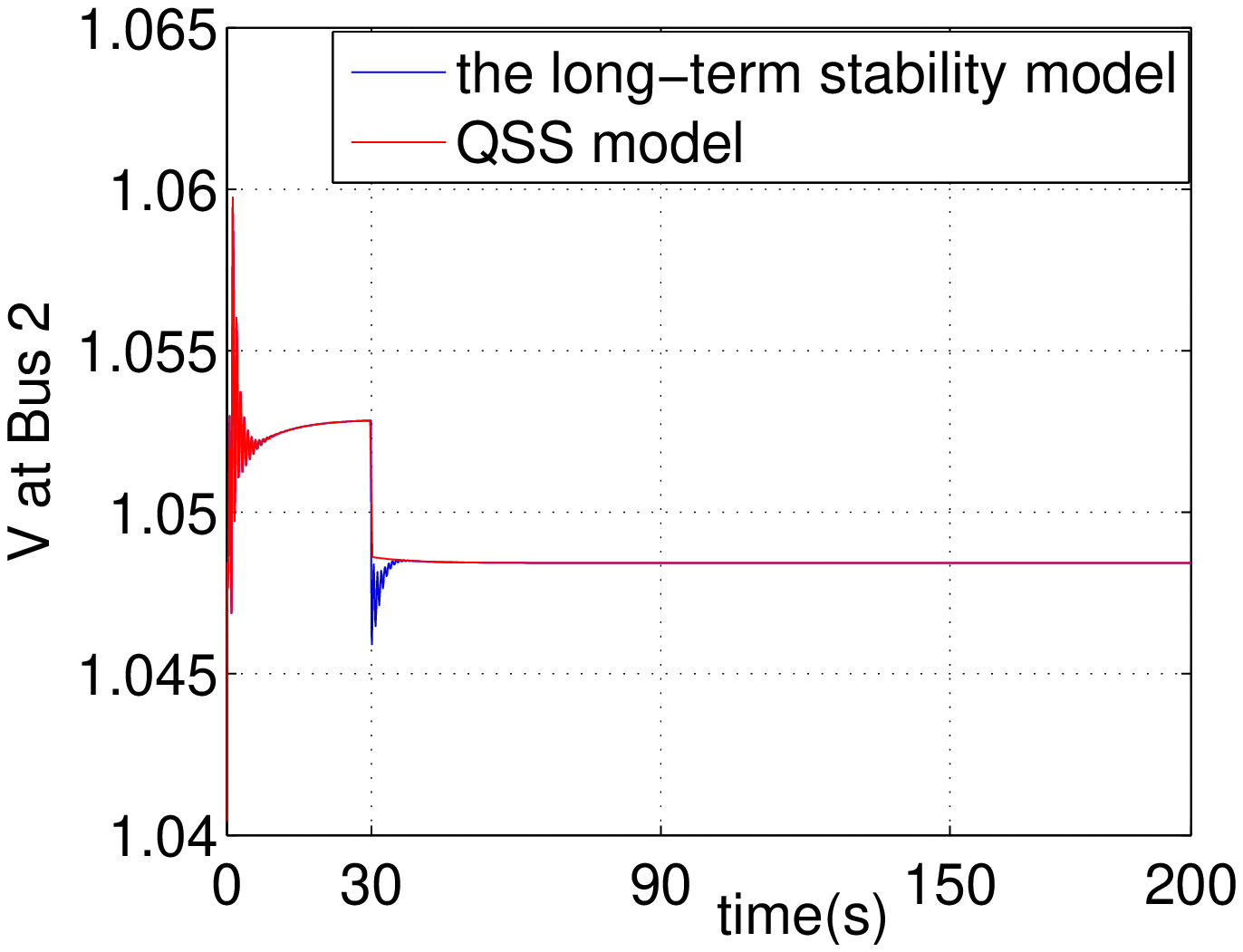}}
\subfloat[]{\label{my14trytransient_31}\includegraphics[width=1.8in ,keepaspectratio=true,angle=0]{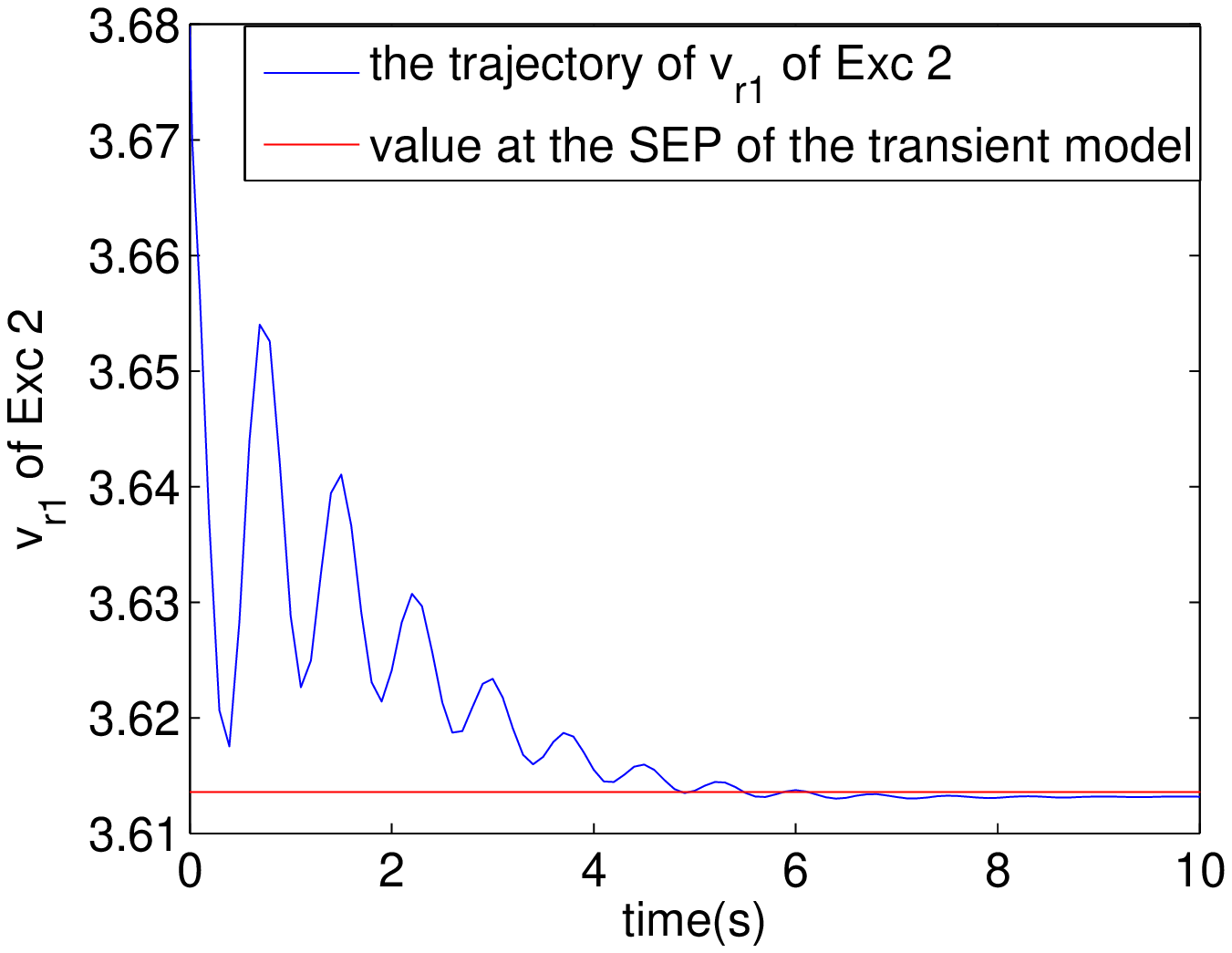}}
\caption{(a). The trajectory comparison of the long-term stability model and the QSS model when $z_d$ were fixed at $z_d(1)$, and both the long-term stability model and the QSS model converged to the same long-term SEP. (b). Trajectory of the transient stability model when $z_d$ changed to $z_d(1)$ at 30s. The trajectory starting from the first point of the long-term stability model converged to the SEP of the transient stability model.}
\end{figure}

\begin{figure}%
\centering
\subfloat[]{\label{vr1vf_trytransient_30}\includegraphics[width=1.8in ,keepaspectratio=true,angle=0]{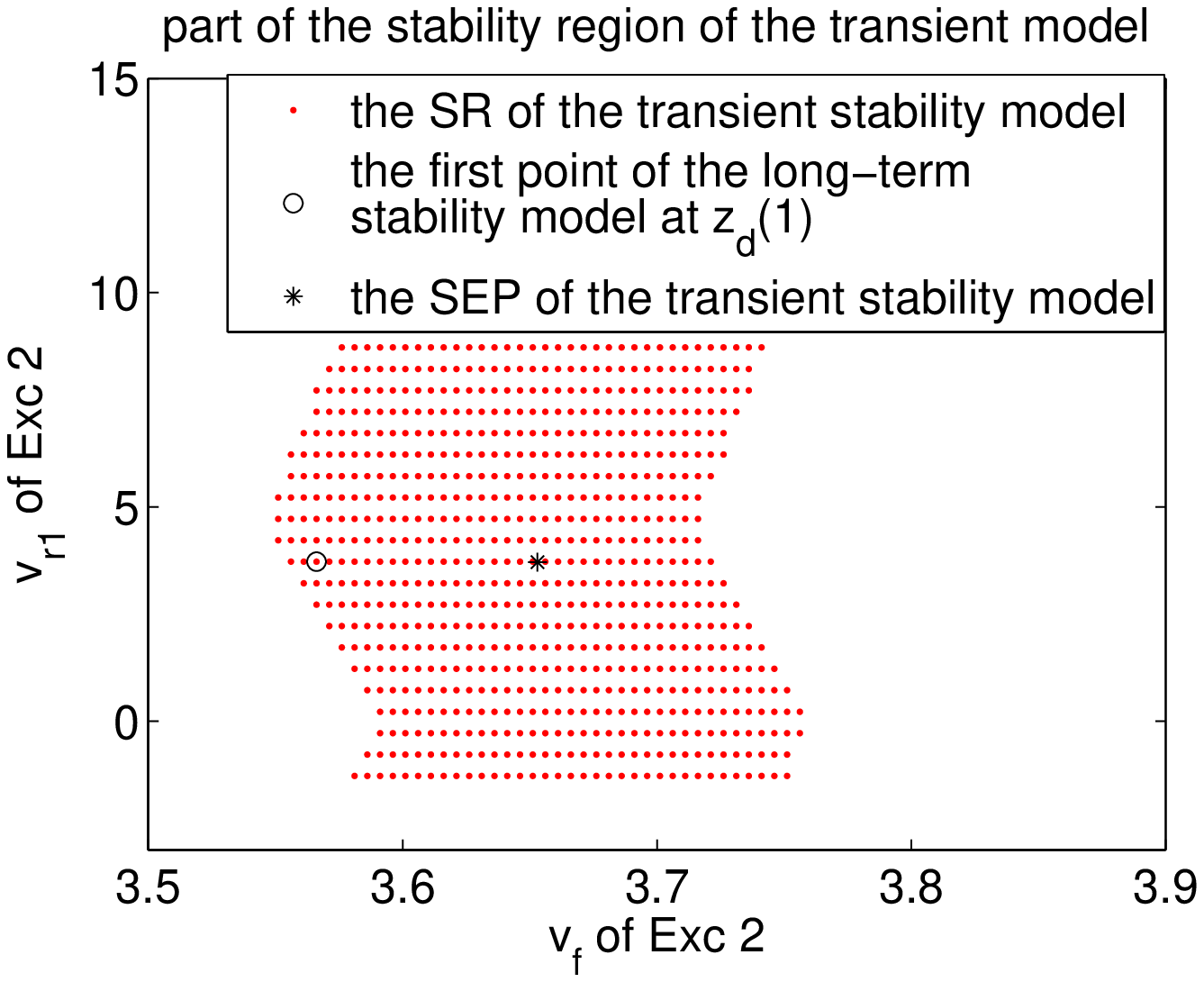}}
\subfloat[]{\label{vr1vf_trytransient_50}\includegraphics[width=1.8in ,keepaspectratio=true,angle=0]{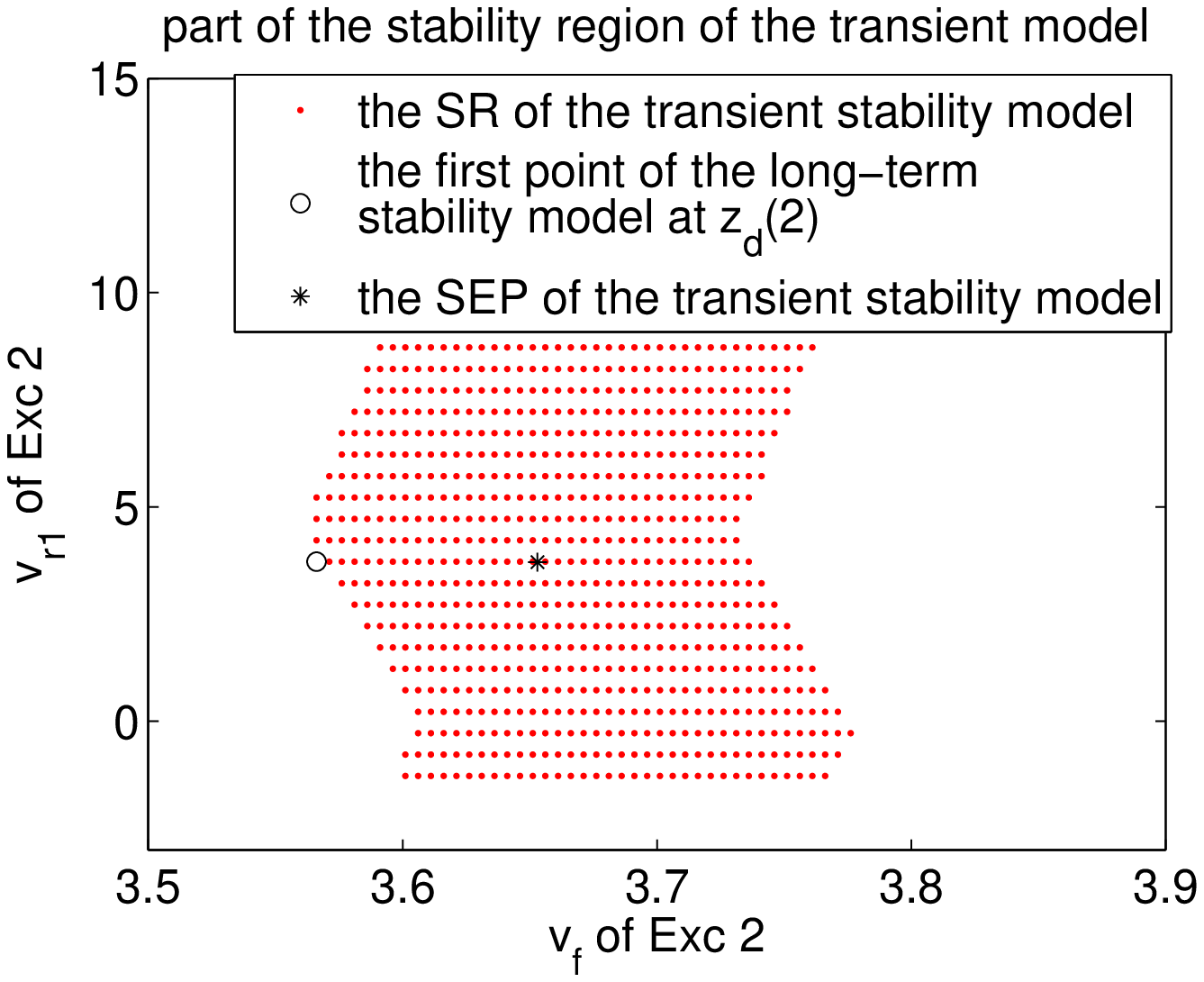}}
\caption{Illustration of Theorems 5 and 6. (a). The stability region of the corresponding transient stability model in the subspace of two fast variables when $z_d$ changed to $z_d(1)$ at 30s. The first point of the long-term stability model was inside the stability region of the corresponding transient stability model; (b). The same as (a) except that $z_d$ changed to $z_d(2)$ at 40s. The first point of the long-term stability model was outside the stability region of the corresponding transient stability model.}
\label{my14try_SR}
\end{figure}

In this case, S1-S3 was also satisfied, and trajectory $\phi_q(\tau,z_{c0},z_d(0),x_0^q,y_0^q)$ of the QSS model moved along $\Gamma_s$. However, condition (b) of Theorem 5 was violated. When $z_d$ jumped from $z_d(0)$ to $z_d(1)$ at 30s, the long-term stability model fixed at $z_d(1)$ was stable, the trajectory comparison is plotted in Fig. \ref{my14try_31} in which both the long-term stability model and the QSS model converged to the same long-term SEP. Fig. \ref{my14trytransient_31} shows the trajectory of a fast variable in the transient stability model and Fig. \ref{vr1vf_trytransient_30} shows stability region of the corresponding transient stability model in the subspace of two fast variables.

\begin{figure}%
\centering
\subfloat[]{\label{my14try_41}\includegraphics[width=1.8in ,keepaspectratio=true,angle=0]{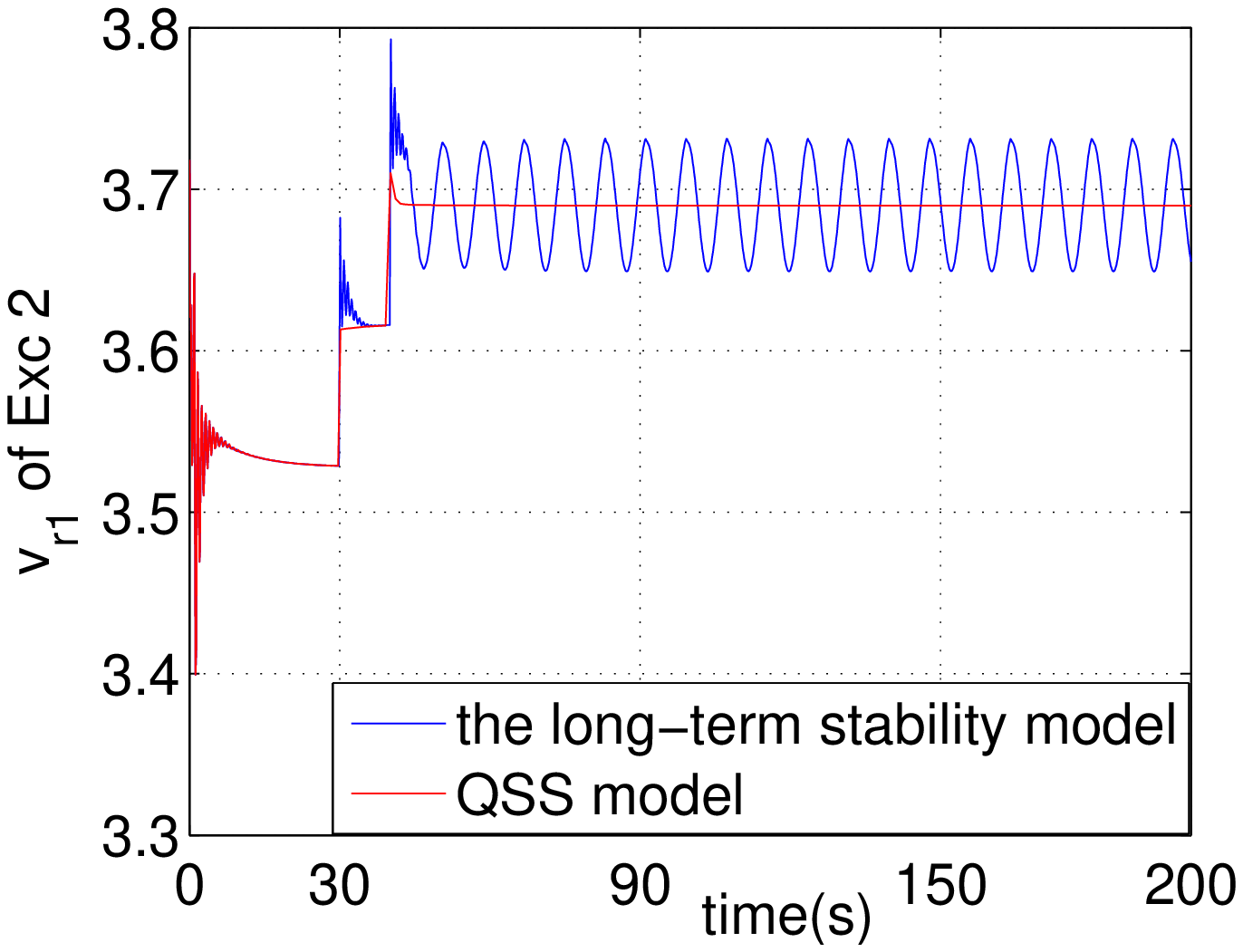}}
\subfloat[]{\label{my14trytransient_41}\includegraphics[width=1.8in ,keepaspectratio=true,angle=0]{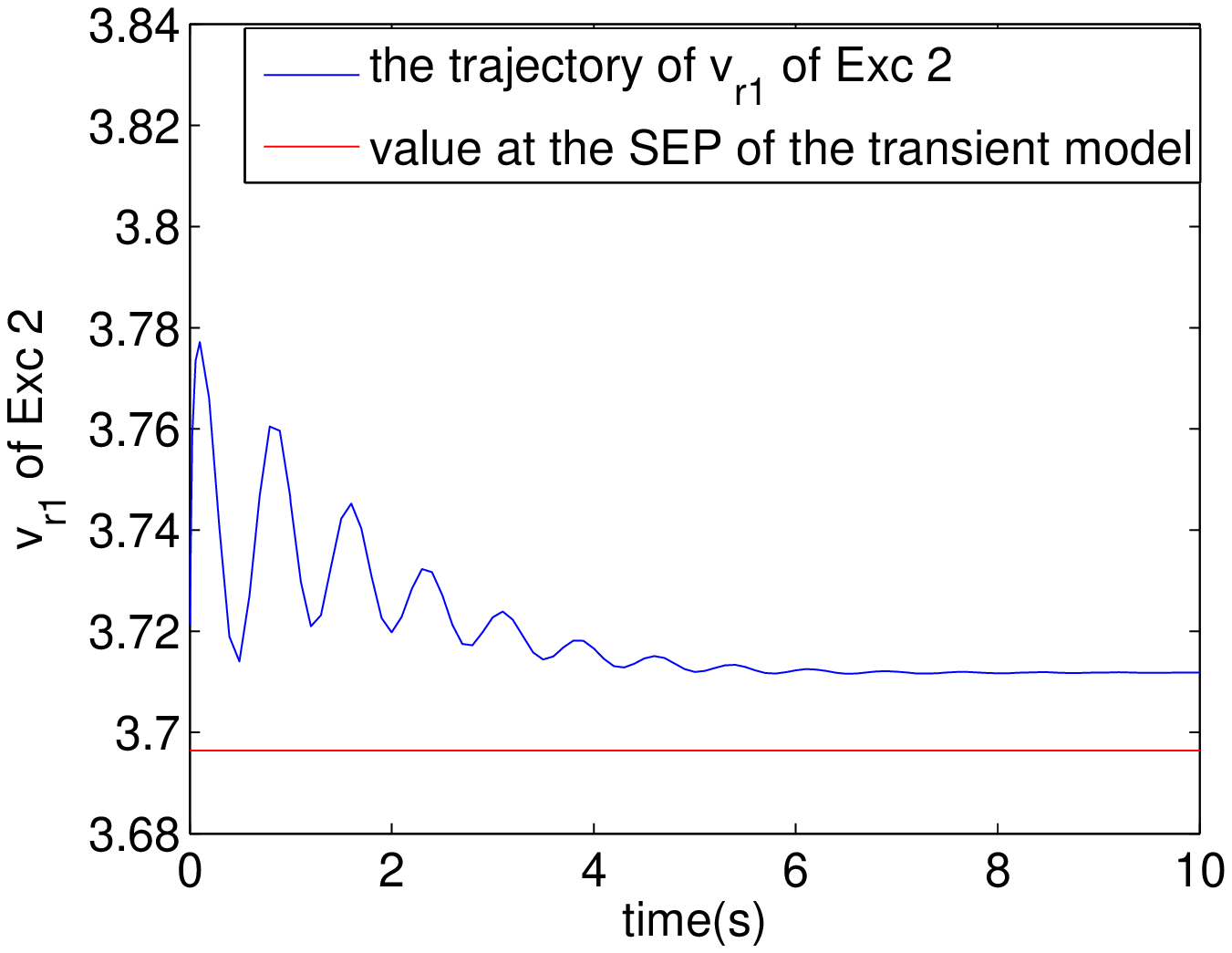}}
\caption{(a). The trajectory comparison of the long-term stability model and the QSS model when $z_d$ were fixed at $z_d(2)$. The long-term stability model became unstable while the QSS model converged to a long-term SEP. (b). Trajectory of the transient stability model when $z_d$ changed to $z_d(2)$ at 40s. The trajectory starting from the first point of the long-term stability model did not converge to the SEP of the corresponding transient stability model.}
\end{figure}

However when $z_d$ jumped from $z_d(1)$ to $z_d(2)$ at 40s, the long-term stability model was no longer stable which can be seen from the trajectory comparison in Fig. \ref{my14try_41}. The transient variables were excited due to the evolution of discrete variables $z_d$ and the trajectory of the long-term stability model was trapped in a stable limit cycle. From a physical viewpoint, the OXL of the generator at Bus 2 reached its limit while the LTC between Bus 2 and Bus 4 tried to restore the voltage at Bus 4 thus required more power support from the generator at Bus 2. The conflict between the OXL and the LTC resulted in the limit cycle shown in Fig. \ref{my14try_41}.

Similarly, Fig. \ref{my14trytransient_41} shows the trajectory of a fast variable in the transient stability model, and Fig. \ref{vr1vf_trytransient_50} shows the stability region of the corresponding transient stability model in the same subspace as Fig. \ref{vr1vf_trytransient_30}. From these two figures it can be seen that the first point $({z}_{c2},z_{d}(2),{x}_2,{y}_2)$ of $\phi_l(\tau,z_{c0},z_d(0),x_0^l,y_0^l)$ after $z_d$ jumped to $z_d(2)$ lied outside the stability region $A_t({z}_{c2},z_{d}(2),l_1({z}_{c2},z_{d}(2)),l_2({z}_{c2},z_{d}(2)))$ of the corresponding transient stability model. As a result, the long-term stability model did not satisfy the condition of consistent attraction.

In summary, fast dynamics were excited by the evolution of long-term discrete dynamics $z_d$ such that the condition of consistent attraction was violated. As a result, the QSS model did not provide correct approximations of the long-term stability model in terms of trajectories and presented incorrect stability assessment in concluding that the long-term stability model was stable while the long-term stability model was long-term unstable.

We provide some physical explanation behind sufficient conditions of the QSS model to explain when the QSS model may fail. In long-term time scale, LTCs are to restore the load-side voltages and hence the corresponding load powers, while OXLs restrict the power support from generators\cite{Cutsem:artical}. The counter effects between LTCs and OXLs further introduce large changes on exciters, leading to long-term instabilities. However, the QSS model assumes that variables of exciters are stable and converge instantaneously fast as LTCs and OXLs evolve, therefore, large changes occurring in the variables of exciters are not reflected in the QSS model. As a result, when the described physical mechanism of long-term instability occurs, the QSS model can fail to provide correct approximations of the long-term stability model.

\section{Conclusions and Perspectives}\label{conclusion}
A theoretical foundation for the QSS model intended for power system long-term stability analysis has been developed. Sufficient conditions for the QSS model to approximate the long-term stability model are derived and relations of trajectory as well as $\omega$-limit point between the long-term stability model and the QSS model are established. Several numerical examples in which the QSS model either succeeds or fails to provide accurate approximations are analyzed using the derived analytical results.

The analytical results derived also point to a research direction for improving the QSS model. It has been shown that the QSS model will provide accurate approximations if the trajectory of QSS model moves along the stable component of the constraint manifold and fast dynamics are not excited by the slow variables. All conditions in Theorem 5 are easy to check except the condition of consistent attraction. If an efficient numerical scheme can be developed to check this condition, then the QSS model can be improved based on the theoretical foundation. It's our intent to develop an improved QSS model to accurately approximate the long-term stability model.

\appendices
\section{Detailed Power System Models\cite{Cutsem:book}\cite{Milano:article}}
\subsection{Generator (GEN):}
Notations are in Table \ref{appedixtable1}.
Dynamic Equations:
\begin{eqnarray}\label{Dfeqn1}
\dot{\delta}&=&\Omega_b(\omega-1)\label{Gen_d1}\\
\dot{\omega}&=&(p_m-p_e-D\omega)/M\\
\dot{e}'_q&=&{(-f_s(e'_q)-(x_d-x'_d){i_d}+v_f^\star)}/{T'_{d0}}\\
\dot{e}'_d&=&{(-e'_d+(x_q-x'_q)i_q)}/{T'_{q0}}\label{Gen_d2}
\end{eqnarray}
$f_s(e'_q)$ is a function for saturation and
\begin{equation}
p_e=(v_q+r_a{i_q})i_q+(v_d+r_a{i_d})i_d
\end{equation}
\begin{equation}
v^{\star}_f=v_f+K_{\omega}(\omega-1)-K_{p}(p-p^0)
\end{equation}
besides $v_p$ and $v_q$ are defined as
$v_d=v\sin(\delta-\theta)$, $v_q=v\cos(\delta-\theta)$,
and following equations describe the relation between the voltage and current:
$0=v_q+r_a{i_q}-e'_q+x'_d{i_d}$,
$0=v_d+r_ai_d-e'_d-x'_q{i_q}$.

Algebraic Equations:
\begin{equation}
\begin{array}{ll}\label{Gen_a1}
0=v_d{i_d}+v_q{i_q}-p&
0=v_q{i_d}-v_d{i_q}-q\\
0=p^0_m-p_m &
0=v^0_f-v_f
\end{array}
\end{equation}

\begin{table}[!ht]
\centering
\caption{Synchronous Machine Variables}\label{appedixtable1}
\begin{tabular}{|c|c|}
\hline
Variable&Description\\
\hline
$\delta$&generator rotor angle\\
\hline
$\omega$&generator rotor speed\\
\hline
$\dot{e}'_q$&q-axis transient voltage\\
\hline
$\dot{e}'_d$&d-axis transient voltage\\
\hline
$p_m$&mechanical power\\
\hline
$p_{m}^0$&initial mechanical power\\
\hline
$v_f$&field voltage\\
\hline
$v_{f}^0$&initial field voltae\\
\hline
$T_{q0}'$&q-axis open circuit transient time constant\\
\hline
$T_{d0}'$&d-axis open circuit transient time constant\\
\hline
$x_q$&q-axis synchronous reactance\\
\hline
$x'_q$&q-axis transient reactance\\
\hline
$r_a$&armature resistance\\
\hline
$M=2H$&mechanical starting time (2$\times$inertia constant)\\
\hline
$D$&damping coefficient\\
\hline
$K_{\omega}$&speed feedback gain\\
\hline
$K_p$&active power feedback gain\\
\hline
$\Omega_b$&base frequency\\
\hline
$p_e$&electrical power\\
\hline
\end{tabular}
\end{table}

\subsection{Automatic Voltage Regulator (AVR):}
Notations are in Table \ref{appedixtable2}.

Dynamic Equations:
\begin{eqnarray}\label{AVR_d1}
\dot{{v}_m}&=&(v-v_m)/T_r\\
\dot{{v}_{r1}}&=&(K_a(v_{ref}-v_m-v_{r2}-\frac{K_f}{T_f}v_f)-v_{r1})/T_a\\
\dot{{v}_{r2}}&=&-(\frac{K_f}{T_f}v_f+v_{r2})/T_f\\\label{Dfeqn2'}
\dot{{v}_{f}}&=&-(v_f(K_e+S_e(v_f))-v_r)/T_e\label{AVR_d2}
\end{eqnarray}
where

\begin{equation}
v_r=\left\{\begin{array}{ll}v_{r1}&\mbox{if }  v_r^{min}\leq{v_{r1}}\leq{v_r^{max}}\\
v_r^{max}&\mbox{if }{v_{r1}}>{v_r^{max}}\\
v_r^{min}&\mbox{if }{v_{r1}}<{v_r^{min}}\\
\end{array}\right.
\end{equation}
and $S_e$ is the ceiling function: $S_e(v_f)=A_ee^{B_e|v_f|}$.

Algebraic Equations:

\begin{eqnarray}
0&=&v_f-v_f^{syn}\label{AVR_a1}\\
0&=&v^0_{ref}-v_{ref}\label{AVR_a2}
\end{eqnarray}

\begin{table}[h]
\centering
\caption{Exciter Variables}\label{appedixtable2}
\begin{tabular}{|c|c|}
\hline
Variable&Description\\
\hline
$v_r^{max}$&maximum regulator voltage\\
\hline
$v_r^{min}$&minimum regulator voltage\\
\hline
$K_a$&amplifier gain\\
\hline
$T_a$&amplifier time constant\\
\hline
$K_f$&stabilizer gain\\
\hline
$T_f$&stabilizer time constant\\
\hline
$K_e$&field circuit integral deviation\\
\hline
$T_e$&field circuit time constant\\
\hline
$T_r$&measurement time constant\\
\hline
$A_e$&$1^{st}$ ceiling coefficient\\
\hline
$B_e$&$2^{nd}$ ceiling coefficient\\
\hline
$v_{ref}(v^0_{ref})$&the reference voltage(or initial)\\
\hline
$v_{r1}$,$v_{r2}$,$v_m$&state variables\\
\hline
\end{tabular}
\end{table}

\subsection{Turbine Governor (TG):}
Notations are in Table \ref{appedixtable4}.

Dynamic Equations:

\begin{eqnarray}\label{Tg_d1}
\dot{x}_{g1}&=&(p_{in}-x_{g1})/T_s\\
\dot{x}_{g2}&=&((1-\frac{T_3}{T_c})x_{g1}-x_{g2})/T_c\\
\dot{{x}}_{g3}&=&((1-\frac{T_4}{T_5})(x_{g2}+\frac{T_3}{T_c}x_{g1})-x_{g3})/T_5\label{Tg_d2}
\end{eqnarray}
where
\begin{eqnarray}
p_{in}^{\star}&=&p_{order}+\frac{1}{R}(\omega_{ref}-\omega)\\
p_{in}&=&\left\{\begin{array}{ll}
p_{in}^{\star}&\mbox{if }p^{min}\leq{p_{in}^{\star}}\leq{p^{max}}\\
p^{max}&\mbox{if }p^{\star}_{in}>p^{max}\\
p^{min}&\mbox{if }p^{\star}_{in}<p^{min}\end{array}\right.\\
p_m&=&x_{g3}+\frac{T_4}{T_5}(x_{g2}+\frac{T_3}{T_c}x_{g1})
\end{eqnarray}

Algebraic Equations:
\begin{equation}\label{Tg_a1}
\begin{array}{ll}
0=p_m-p_m^{syn}&
0=\omega_{ref}^{0}-\omega_{ref} 
\end{array}
\end{equation}

\begin{table}[h]
\centering
\caption{Turbine Governor Variables}\label{appedixtable4}
\begin{tabular}{|c|c|}
\hline
Variable&Description\\
\hline
$\omega_{ref}^{0}$&reference speed\\
\hline
$R$&droop\\
\hline
$p^{max}$&maximum turbine output\\
\hline
$p^{min}$&minimum turbine output\\
\hline
$T_s$&governor time constant\\
\hline
$T_c$&servo time constant\\
\hline
$T_3$&transient gain time constant\\
\hline
$T_4$&power fraction time constant\\
\hline
$T_5$&reheat time constant\\
\hline
$x_{gi}$&state variables (i=1,2,3)\\
\hline
\end{tabular}
\end{table}

\subsection{Over Excitation Limiter (OXL):}
Notations are in table \ref{appedixtable5}.

Dynamic Equations:
\begin{eqnarray}\label{OXL_d1}
{\dot{v}_{OXL}}&=&\begin{array}{ll}(i_f-i_f^{lim})/T_0& \mbox{ if }i_f>i_f^{OXL}\end{array}\\\nonumber
{\dot{v}_{OXL}}&=&\begin{array}{ll}0&\qquad\qquad\qquad\mbox{if } i_f\leq{i_f^{OXL}}\end{array}
\end{eqnarray}

Algebraic Equations:
\begin{eqnarray}\label{OXL_a1}
0&=&\sqrt{(v+\gamma_q)+p^2}+(\frac{x_d}{x_q}+1)\\
&&\frac{\gamma_q(v+\gamma_q)+\gamma_p}{\sqrt{(v_g+\gamma_q)^2+p^2}}-i_f\nonumber\\
0&=&{v^0_{ref}}-v_{ref}+v_{OXl}\label{OXL_a2}
\end{eqnarray}
with $\gamma_p={x_q}p/v$, $\gamma_q={x_q}q/v$. And the over excitation limiter starts to work after a fixed delay $T_0$ regardless of the field current overload.

\begin{table}
\centering
\caption{Over Excitation Limiter Variables}\label{appedixtable5}
\begin{tabular}{|c|c|}
\hline
Variable&Description\\
\hline
$x_d$&d-axis estimated generator reactance\\
\hline
$x_q$&q-axis estimated generator reactance\\
\hline
$i_f$&synchronous machine field current\\
\hline
$i_f^{lim}$&maximum field current\\
\hline
$v^0_{ref}$&the reference voltage of automatic voltage regulator\\
\hline
$T_0$&integrator time constant\\
\hline
$p(\mbox{or }q)$&active (or reactive) power of generator\\
\hline
$K_0$&fixed time delay\\
\hline
$v_{OXL}$&state variabe\\
\hline
\end{tabular}
\end{table}

\subsection{Exponential Recovery Load (ERL):}
Notations are in table \ref{appedixtable6}.

Dynamic Equations:
\begin{eqnarray}\label{ERL_d1}
{\dot{x}_p}&=&-x_p/T_p+p_s-p_t\\
{\dot{x}_q}&=&-x_q/T_q+q_s-q_t\label{ERL_d2}
\end{eqnarray}
where $p_s$ and $p_t$ are the static and transient real power absorptions, similar definition for $q_s$ and $q_t$. $p^0_L$ and $q^0_L$ are PQ load power from power flow solutions. Besides,
$p^0=\frac{k_p}{100}p^0_L$, $q^0=\frac{k_q}{100}q^0_L$, $p_s=p^0(v/v^0)^{\alpha_s}$, $p_t=p^0(v/v^0)^{\alpha_t}$, $q_s=p^0(v/v^0)^{\beta_s}$, $q_t=p^0(v/v^0)^{\beta_t}$.

Algebraic Equations:
\begin{eqnarray}\label{ERL_a1}
p&=&x_p/T_p+p_t\\
q&=&x_q/T_q+q_t\label{ERL_a2}
\end{eqnarray}

\begin{table}[h]
\centering
\caption{Exponential Recovery Load Variables}{\label{appedixtable6}}
\begin{tabular}{|c|c|}
\hline
Variable&Description\\
\hline
$k_p$&active power percentage\\
\hline
$k_q$&reactive power percentage\\
\hline
$T_p$&active power time constant\\
\hline
$T_q$&reactive power time constant\\
\hline
$\alpha_s$&static active power exponent\\
\hline
$\alpha_t$&dynamic active power exponent\\
\hline
$\beta_s$&static reactive power exponent\\
\hline
$\beta_t$&dynamic reactive power exponent\\
\hline
$x_p$,$x_q$&state variabels\\
\hline
\end{tabular}
\end{table}
\subsection{Load Tap Changer (LTC):}
\begin{equation}\label{LTC_d1}
m_{k+1}=\left\{\begin{array}{ll}m_k+\triangle{m}&\mbox{if  }v>v_0+d\mbox{  and  }m_k<m^{max}\\
m_k-\triangle{m}&\mbox{if  }v<v_0+d\mbox{  and  }m_k>m^{min}\\
m_k &\mbox{otherwise}\end{array}\right.
\end{equation}
The tapping delay are assumed to be independent of $V$, but larger for first tap change than for the subsequent ones while without the inverse time characteristic. Refer to \cite{Cutsem:book} for more details.

\section{Detailed and Generic long-term stability models}
The detailed and generic long-term stability model are shown in Table \ref{detailed&generic}. Moreover, the detailed variables and their corresponding generic variables $z_c$, $z_d$, $x$ and $y$ are also indicated.

\begin{table*}
\caption{long-term stability model and corresponding generic variables}
\begin{center}\normalsize
\begin{tabular}{|*{7}{l|}}
\hline
Detailed Long-Term Stability Model&Generic Long-Term Stability Model&Detailed Variables\\
\hline
TG: (\ref{Tg_d1})-(\ref{Tg_d2}), OXL: (\ref{OXL_d1}),&$\dot{z}_c=\ee h_c(z_c,z_d,x,y)$&slow continuous variables $z_c$:\\
ERL: (\ref{ERL_d1})-(\ref{ERL_d2}).&&TG: $x_{g1},x_{g2},x_{g3}$, OXL: $v_{OXL}$, ERL: $x_p$,$x_q$.\\
\hline
LTC: (\ref{LTC_d1}).&$z_d(k+1)=h_d(z_c,z_d(k),x,y)$&slow discrete variables $z_d$: $m_k$.\\
\hline
GEN: (\ref{Gen_d1})-(\ref{Gen_d2}),&$\dot{x}=f(z_c,z_d,x,y)$&fast continuous variables $x$:\\
AVR: (\ref{AVR_d1})-(\ref{AVR_d2}).&&GEN: $\delta$,$\omega$,$e'_q$,$e'_d$, AVR: $v_m$,$v_{r1}$,$v_{r2}$,$v_f$.\\
\hline
TG:(\ref{Tg_a1}), OXL:(\ref{OXL_a1})-(\ref{OXL_a2}), &$0=g(z_c,z_d,x,y)$&algebraic variables $y$:\\
ERL:(\ref{ERL_a1})-(\ref{ERL_a2}), GEN:(\ref{Gen_a1}),&&TG: $\omega_{ref}$, OXL: $i_f$, AVR: $v_{ref}$,\\
AVR:(\ref{AVR_a1})-(\ref{AVR_a2}), power relations.&&GEN: $p$,$q$,$p_m$,$v_f$, Bus: $v$ and $\theta$.\\
\hline
\end{tabular}
\end{center}
\label{detailed&generic}
\end{table*}

\section*{Acknowledgment}
The authors would like to thank Prof. Luis F. C. Alberto and Dr. Tao Wang for helpful discussions.
\ifCLASSOPTIONcaptionsoff
  \newpage
\fi

\begin{IEEEbiography}[{\includegraphics[width=1in,height=1.25in,clip,keepaspectratio,angle=0]{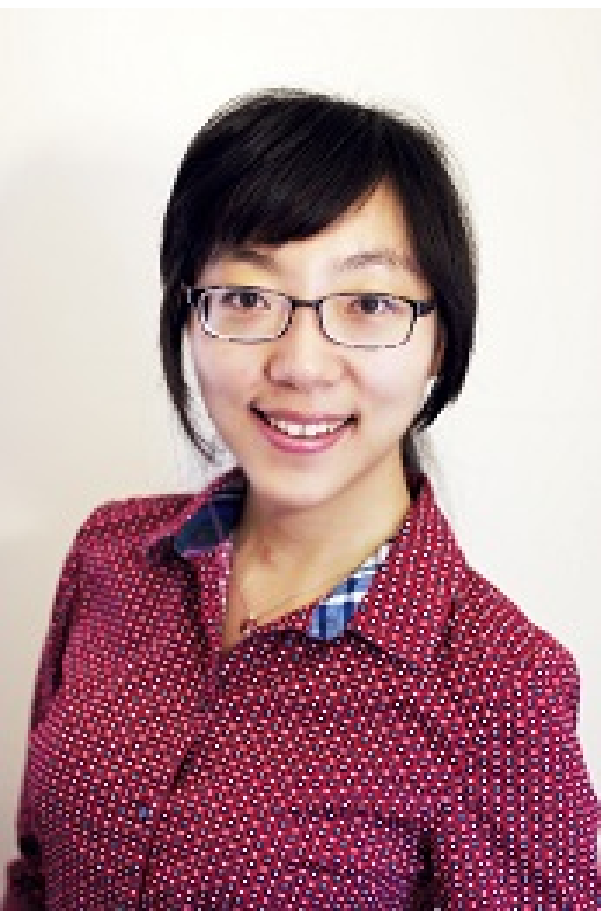}}]{Xiaozhe Wang}
received the B.S. degree in Information Science \& Electronic Engineering from Zhejiang University, Zhejiang, China, in 2010, M.Eng degree in Electrical and Computer Engineering from Cornell University, Ithaca, NY in 2011. She is currently pursuing the Ph.D. degree in the same school. She was a Fung scholar in 2008-2009, and a Jacobs Scholar in 2011-2012. Her research interests include nonlinear systems, power system stabilities, and optimization in power systems.
\end{IEEEbiography}

\begin{IEEEbiography}[{\includegraphics[width=1in,height=1.25in,clip,keepaspectratio,angle=0]{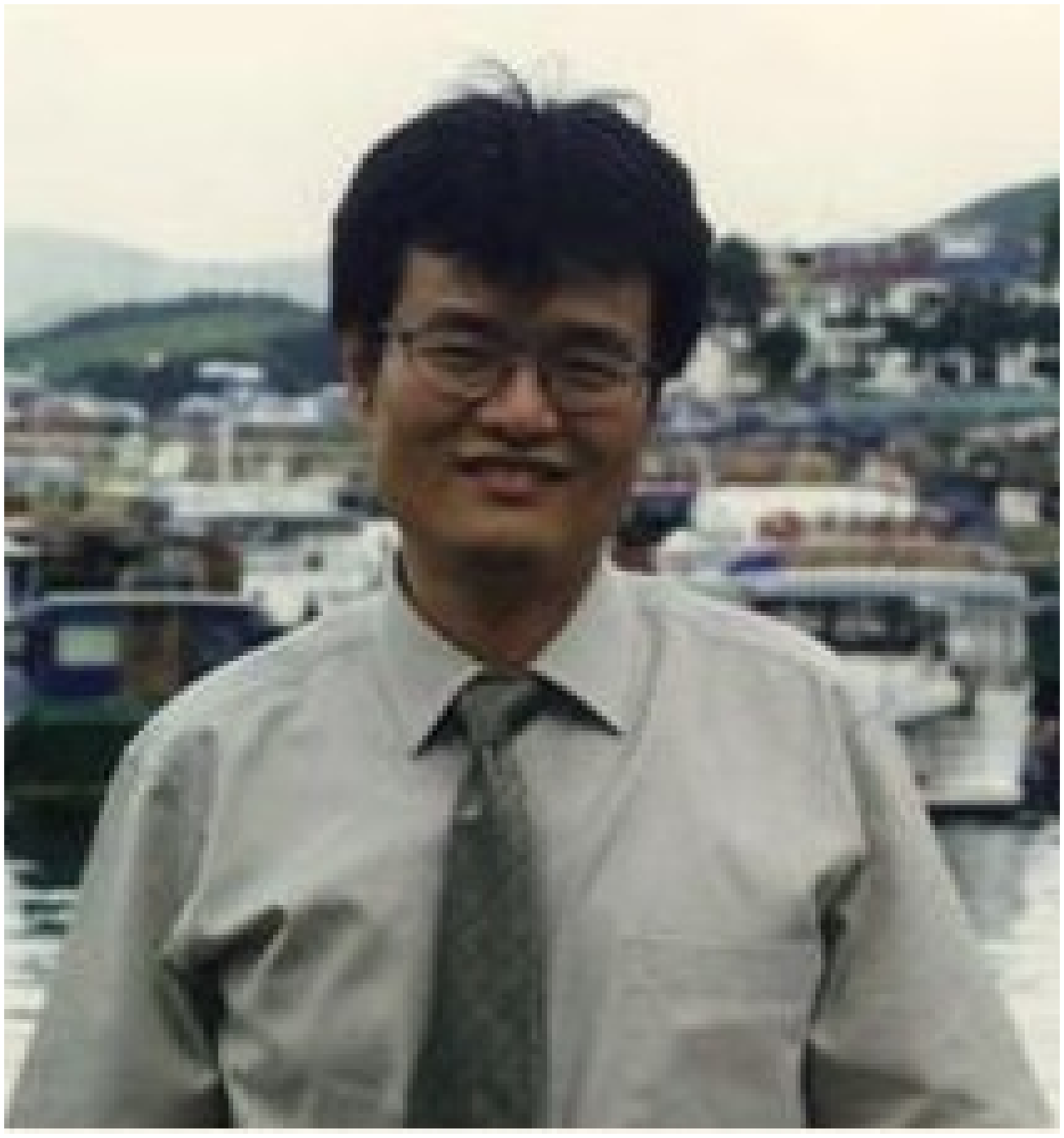}}]{Hsiao-Dong Chiang}
(M'87-SM'91-F'97) received the PhD degree in EECS from University of California, Berkeley in 1986 and is currently Professor of ECE at Cornell University, Ithaca, New York. His research effort is focused on theoretical developments and practical applications in the areas of nonlinear system stability and control theory, nonlinear computation, and nonlinear optimization methods. He and his colleagues have published more than three hundred and fifty referred papers and have been awarded 14 patents arising from their research and development work both in the United States and internationally. He was Associate Editor of the IEEE Trans. CAS from 1990 to 1991, and was Associate Editor for Express Letters of the IEEE Transactions on CAS Part I from 1993 to 1995. He is currently on the Editorial board of two international Journals. He is Author of the book "Direct Methods for Power System Stability Analysis: Theoretical Foundation, BCU Methodology and Applications", John Wiley \& Sons, 2010.
\end{IEEEbiography}

\end{document}